\documentclass[prl,twocolumn]{revtex4-1}
\usepackage{amsmath,amssymb}
\usepackage{graphicx}
\usepackage{mathrsfs}

\def\unit#1{\ensuremath{\mathrm{\,#1}}}
\newcommand{\I}{{\rm i}}
\newcommand{\D}{\mathrm{d}}
\newcommand{\E}{\mathrm{e}}
\newcommand{\avg}[1]{\ensuremath{\left\langle #1 \right\rangle}}

\begin{document}

\title{Optical correlation techniques \\for the investigation of colloidal systems}
\author{Roberto Piazza}
\affiliation{Dipartimento di Chimica, Materiali e Ingegneria
Chimica ``Giulio Natta'', Politecnico di Milano, Milano, Italy}
%\date{}
\begin{abstract}
This review aims to provide a simple introduction to the
application of optical correlation methods in colloidal science.
In particular, I plan to show that full appraisal of the intimate
relation between light scattering and microscopy allows designing
novel powerful investigation techniques that combine their powers.
An extended version of this paper will appear in \emph{Colloidal
Foundations of Nanoscience}, edited by D. Berti and G. Palazzo,
Elsevier (ISBN 978-0-444-59541-6). I am very grateful to the
publisher for having granted me the permission to post this
preprint on arXiv.
\end{abstract}
 \maketitle

Scattering or microscopy experiments necessarily involve
statistical fluctuations, which already stem from the optical
source used to probe the investigated system, are modified by the
interaction of the probing field with the sample, and are further
influenced by the detection process. All these effects concur in
turning optical fields into random signals, which are physically
described in terms of correlations. In optics, fluctuations and
correlations are nicely embodied  in the concept of
\emph{coherence}. As the Roman god of beginnings and transitions,
Janus, coherence is however two-faced: because the field
fluctuates both in time and space, one should indeed distinguish
between \emph{temporal} and \emph{spatial} coherence. Setting
apart these two aspects is not always possible, since they can be
intrinsically intermixed, but when this is feasible, it is far
more than a useful practical approach. As a matter of fact, it
involves an important conceptual distinction: whereas temporal
coherence is a \emph{physical} concept, related to the spectrum of
the optical signal generated by the interaction of the incoming
field with the sample, which is therefore the actual ``source'' of
the detected radiation, spatial coherence has mostly to do with
the source extension, so it is usually (but not always) a
\emph{geometrical} problem. Curiously, in spite of this, spatial
coherence is far more important, for the physical problems we
shall investigate, than temporal coherence. Nevertheless, it is
useful to start by recalling some basic concepts of the latter. We
shall first refer to the temporal coherence properties of optical
fields, or ``first order'' optical coherence, to distinguish it
from correlations of the \emph{intensity}, discussed later.

\section{Basic concepts in statistical optics}
\label{s.statopt}
\subsection{Temporal coherence} \label{s.temcoh}   Temporal fluctuations can be
equivalently discussed in the frequency domain, where it is
basically related to non-monochromaticity. For a generically
time--varying real field $u^R(t)$ with Fourier transform
$\mathscr{F}[u^R]= \tilde{u}^R(\omega)$, it is useful to introduce
the associated \emph{analytical signal}\cite{GoodmanSO}
\begin{equation}\label{anasig}
    u(t) =\frac{1}{\pi}\int_0^\infty \D \omega\,
    \tilde{u}^R(\omega) \E^{-\I \omega t} = 2\int_0^\infty \D
    \nu\,
    \tilde{u}^R(\nu)\E^{-\I 2\pi \nu},
\end{equation}
which is then a complex quantity obtained by suppressing the
negative frequency components of $u^R (t)$ and doubling the
amplitude of the positive ones.\footnote{This is nothing but an
extension of what is done in representing a monochromatic signal
$u^R(t) =A\cos(\omega_0 t -\phi)$ as  $u(t) = A\exp[-\I( \omega_0
t -\phi)]$, as can be appreciated by looking at the Fourier
transform (FT) in time of these two functions:
$$ \mathscr{F}[u^R] = (A/2) [ \E^{\I\phi} \delta(\omega - \omega_0)+ \E^{-\I\phi} \delta(\omega + \omega_0)]
\;\;;\;\; \mathscr{F}[u] = A\E^{\I\phi} \delta(\omega -
\omega_0),$$}   For a \emph{narrowband} signal, having a spectrum
centered on $\omega_0$ of width \mbox{$\Delta \omega \ll
\omega_0$}, we can write $u^R(t) = A(t) \cos[\omega_0 t
-\phi(t)]$, hence $u(t) = U(t)\E^{-\I\omega_0 t}$, where $U(t)
=A(t)\E^{\I \phi(t)}$ is called the \emph{complex envelope}.

The crucial point that we are going to discuss is that any signal
with finite bandwidth \emph{must} display temporal fluctuations:
specifically, the envelope $U(t)$ of a signal with bandwidth
$\Delta \omega$ does not appreciably change in time on time scales
much shorter than a \emph{coherence time} $\tau_c = 2\pi / \Delta
\omega$, to which we can associate a coherence length $\ell_c
=c\tau_c$. To see this, let us introduce the time correlation
function of the analytic signal, or \emph{self-coherence function}
\begin{equation}\label{self-coh}
    \Gamma(\tau) = \avg{u^*(t)u(t+\tau)}_t,
\end{equation}
where the average is performed over the initial time $t$, and we
assume the process to be stationary, so that $\Gamma$ does not
depend explicitly on $t$. Normalizing $\Gamma(\tau)$ to is initial
value $\Gamma(0) = \avg{|u(t)|^2}_t = I$, we obtain the
\emph{degree of first order coherence} (usually simply dubbed
``field correlation function'')
\begin{equation}\label{g1}
    g_1(\tau) =\frac{\avg{u^*(t)u(t+\tau)}_t}{I}
\end{equation}
Provided that a signal has a finite average power we can define
its  \emph{power spectral density}
\begin{equation}\label{power:spectrum}
    P^R_u(\omega) = \lim_{T\rightarrow \infty}\frac{1}{T}\int_{-T}^T \D t \, u^R(t)\E^{\I\omega
t}
\end{equation}
From the definition~(\ref{anasig}) it can be easily shown that the
power spectrum $P_u(\omega)$ of the complex analytic signal is
just $4P_u^R(\omega)$ for $\omega\ge 0$, and 0 otherwise. The
fundamental link between the time and frequency description is
then provided by the \emph{Wiener-Kintchine} (WK) \emph{theorem},
which states that $\Gamma(\tau)$ and $P_u(\omega)$ are
\emph{Fourier transform pairs}. If we \emph{define} the normalized
power spectrum of the real signal as
\begin{equation}\label{normalized PS}
   P(\omega) =\left\{\begin{array}{cc}
                       \dfrac{P_u^R(\omega)}{\int_0^\infty \D\omega\,P_u^R(\omega)} & \mathrm{for}~\omega\ge 0 \\
                       0 & \mathrm{for}~\omega < 0 \\
                     \end{array}\right.
\end{equation}
the WK theorem can be restated in the form
\begin{equation}\label{WK}
    \left\{\begin{array}{l}
             P(\omega) = {F}[g_1(\tau)] = \int_{-\infty}^\infty \D\tau g_1(\tau) \E^{\I\omega\tau} \vspace{5pt}\\
             g_1(\tau)= \mathscr{F}^{-1}[P(\omega)] = \int_{-\infty}^\infty \D\omega P(\omega) \E^{-\I\omega\tau}, \\
           \end{array}\right.
\end{equation}
which will be particularly useful for our purposes. The degree of
temporal coherence is strongly related to the signals detected in
classical interferometric measurements, such as those obtained
with a Michelson interferometer.\cite{Loudon} Qualitatively, the
beams propagating in the two arms of the interferometer can
interfere only if the difference $\Delta l$ between the optical
paths is smaller than the coherence length of the source $\ell_c$.
Quantitatively, one finds that the time dependence of the detected
intensity is given by
\begin{equation}\label{interf}
   I = I_0 \{1+\mathrm{Re}[g_1(\Delta t)]\},
\end{equation}
with $\Delta t = \Delta l/c$, which is then proportional to the
\emph{real} part of the time correlation function, evaluated at
the delay $\Delta t$.

As an important example for what follows,  we briefly describe the
temporal properties of a \emph{narrowband thermal source}, defined
as a collection of many microscopic independent emitters, such as
a collection of thermally excited atoms, all radiating at the same
frequency $\omega_0$, but undergoing collisions that induce abrupt
phase jumps. With $N$ identical emitters, the total signal
amplitude (the complex envelope) can be written
$$U(t) = A(t)\E^{\I\phi(t)}= \sum_{i=1}^N u_i(t) =
a\sum_{i=1}^N\E^{\I\phi_i(t)},$$ where $u_i(t)= a\E^{\I\phi_i(t)}$
is the complex envelope for a single emitter. This is nothing but
a $N$-step random walk in the complex plane. For large $N$, $u_r =
\mathrm{Re} (U) = A \cos(\phi)$ and $u_i = \mathrm{Im} (U)= A
\sin(\phi)$ have therefore a \emph{joint Gaussian statistics}
\begin{equation}\label{field statistics}
   p(r,i) =
   \frac{1}{2\pi\sigma^2}\exp\left(-\frac{u_r^2+u_i^2}{2\sigma^2}\right),
\end{equation}
with $\sigma = a\sqrt{N}$. By a standard transformation of
variables, it is  easy to show the probability density for the
amplitude is a Rayleigh distribution
\begin{equation*}\label{rayleigh}
    p_A(A) = \frac{A}{\sigma^2}\exp\left(-\frac{A^2}{2\sigma^2}\right) \,\,\, (A\ge 0)
\end{equation*}
A photodetector does not respond to the \emph{instantaneous}
optical intensity associated to the signal, but rather to its
value averaged over many optical cycles that, for a narrowband
signal, is $I_{rad}=(\epsilon_0 c/2)A^2$, where $\epsilon_0$ is
the vacuum permittivity and $c$ the speed of light. Following a
common convention, rather than the ``radiometric'' intensity
$I_{rad}$, we shall simply call ``intensity'' the quantity $I=
A^2$ (actually an \emph{irradiance}). Changing again variable, we
get
\begin{equation}\label{Intensity statistics}
   P_I(I) =
   \frac{1}{2\sigma^2}\exp\left(-\frac{I}{2\sigma^2}\right)=
    \frac{1}{\avg{I}}\exp\left(-\frac{I}{\avg{I}}\right).
\end{equation}
The intensity has therefore an \emph{exponential} probability
density, with a decay constant given by its average value
$\avg{I}$.

These probability distributions for the field and intensity apply
for instance to a spectral lamp, but also, as we shall see, to a
medium containing scatterers. As a matter of fact, a gaussian
distribution for the field characterizes any ``random'' optical
source. However, the spectrum and the time--correlation function
depend on the \emph{physical origin} of the frequency broadening.
Indeed, for independent emitters, we have $\avg{u_i(0)u_j(t)} =0$
for $i\ne j$. Hence:
$$ \Gamma(\tau) = \avg{U^*(0)U(\tau)}= \sum_{i=1}^N \avg{u_i(0)u_i(\tau)} =
N\avg{u(0)u(\tau)}.$$ The field correlation function of the system
coincides therefore with the correlation function for a
\emph{single} emitter, $g_1(\tau) \equiv g_1^{(i)}(\tau)$, which
is determined by a specific physical mechanism. Let us for
instance consider the model we formerly introduced, corresponding
to a ``collision-broadened'' source, where  $g_1(\tau)
=\E^{-\I\omega_0 \tau}\avg{\E^{\I[\phi(\tau)-\phi(0)]}}.$ The
phases $\phi(0)$ and $\phi(t)$ are correlated only if the atom
does not undergo collisions in $\tau$, so the phase correlation
function is proportional to the probability of colliding at any
$t> \tau$, which is easily found to be $\exp(-\tau/\tau_c)$, where
$\tau_c$ is the average time between collisions. Hence
\begin{equation}\label{correl_thermal}
   g_1(\tau) = \exp(-\I\omega_0 \tau -\tau/\tau_c),
\end{equation}
with $\tau_c$ playing therefore the role of coherence time (for a
gas at $300\unit{K}$, $10^5\unit{Pa}$, $\tau_c \simeq 30 ps$ and
$\ell_c \simeq 1\unit{cm}$). It is for instance easy to show that,
in a Michelson interferometer, the fringe visibility is related to
$\tau_c$ by $$\frac{I_{max} -I_{min}}{I_{max} +I_{min}}=
\E^{-\Delta t/\tau_c},$$ where $\Delta t$ is the difference in
propagation time between the two arms. Fourier--transforming
$g_1(t)$, we obtain a Lorenzian lineshape for the power spectrum
\begin{equation}\label{spectrum_thermal}
    P(\omega) = \frac{1}{\pi\tau}\frac{1}{(\omega -\omega_0)^2+(1/\tau)^2}
\end{equation}

In view of our application to light scattering, it is also useful
to have a brief look to the temporal coherence of a laser source.
Even when operating on a single longitudinal mode, like the
diode--pumped solid--state lasers (DPSS) now extensively used in
light scattering measurements, a laser is \emph{not} an ideal
monochromatic source, for is displays phase fluctuations due to
the intrinsic nature of the lasing process but also, in practice,
to coupling with mechanical vibrations of the cavity mirrors. Well
above lasing threshold and at steady--state, the field amplitude
can be written as\cite{Armstrong}
$$ u^R(t) = A\cos[\omega_0\vartheta(t)]+u_n(t).$$
where $u_n(t)$ is a narrowband noise due to spontaneous emission,
while phase fluctuations are embodied in $\vartheta(t)$.
Neglecting the additive noise contribution, which is usually very
small, neither the amplitude nor the intensity probability
densities differ however from those of an ideal monochromatic
source. Mechanical stability usually sets a lower limit of the
order of tens of MHz to the laser bandwidth, which is  far wider
than the extremely narrow line of an ideal single--mode laser:
yet, this is mostly due to phase fluctuations, hence intensity
fluctuations are usually negligible. However, scattering
measurements are often still made using common lab sources, such
as simple He-Ne lasers, which oscillates on \emph{many}
longitudinal modes separated by $c/2L$, where $L$ is the cavity
length. By increasing the number of oscillating modes, and
provided that coupling between different modes is weak, the
intensity fluctuations approach those of a thermal source with a
bandwidth equal to that of the  atomic gain line of the laser.

\subsection{Spatial coherence}\label{s.spacoh}
Suppose we illuminate with a laser beam a light diffuser, for
instance a window made of ground glass: then, a complex figure
made of many irregular spots forms on a screen placed beyond the
diffuser, which is what we call a \emph{speckle pattern}. If we
insert a lens and enlarge the beam spot on the diffuser, the
speckle size reduces. Conversely, if we move the diffuser towards
the lens focus plane, the speckle pattern becomes much coarser.
Hence, the speckle size depends on the extension of the
illuminated region on the diffuser.

Again, reflecting upon  an interferometric experiment, in this
case made with a classical two-pinhole Young's setup, sheds light
on the origin of this effect. When an absorbing screen pierced by
two pinholes $P_1$ and $P_2$ separated by a distance $d$ is
illuminated by a monochromatic \emph{point-like} source, fringes
with a spatial period $\Delta x = l\lambda/d$ form on a plane
placed at distance $l$ from the screen. However, if we illuminate
the pinholes with an \emph{extended} source $\Sigma$ of size $D$
made of independent emitters and placed at distance $z$ from the
screen, the fringe pattern forms only provided that $Dd/z \ll
\lambda$.

Fringe visibility is actually a manifestation of the \emph{spatial
coherence} of the fields at the pinholes. Consider indeed two
points $U$ and $V$ on $\Sigma$, which we assume to be a thermal
source made of many independent and spatially uncorrelated
emitters, and call $u_i$ and $v_i$ the amplitude of the fields
reaching pinhole $P_i$ from $U$ and $V$ respectively. If $P_1$ and
$P_2$ are very close, so that $u_1\simeq u_2$, $v_1\simeq v_2$),
the fields $U(P_1) = u_1 +v_1$, $U(P_2) = u_2 +v_2$, will be
\emph{strongly} correlated (they are almost the same field!), even
if the fields $u$ and $v$ are fully uncorrelated. Namely,
propagation from $\Sigma$ to the screen \emph{induces} spatial
correlations even if different points of the source are
uncorrelated.

However, if $P_2$ is moved apart from $P_1$, the phases of the
fields coming from $U$ and $V$ change differently. If $r_u(P_i)$
and $r_v(P_i)$ are the distances of $U$ and $V$ from pinhole
$P_i$, putting $\Delta r_u = r_u(P_1) -r_u(P_2)$, $\Delta r_v =
r_v(P_1) -r_v(P_2)$ we have at first order $\Delta r_u = - \Delta
r_v \simeq dD/z$, where $D$ is the distance $\overline{UV}$.
Spatial field correlation is retained only provided that $\Delta
r_u - \Delta r_v \ll \lambda$, namely, $d \ll \lambda z/D$. In the
Young setup, the fields coming from $U$ and $V$ form two displaced
sets of fringes. However, if the pinhole are sufficiently close,
fringe oscillations are coarse, the shift of the two patterns is a
small fraction of their period, and the sum of the two
interference patterns still shows fringes. Conversely, if the
pinholes are moved apart, fringe oscillation becomes more rapid
and the two sets of fringes soon gets strongly out of phase,
canceling out.

When $U$ and $V$ are taken as far as possible, so that $D$ is the
maximal lateral extension of the source, the pinholes must
therefore lie within a \emph{coherence area} $A_c \simeq
(z\lambda/D)^2$. To the source is then associated a ``coherence
cone'' with solid angle at vertex $\Delta \Omega \simeq
(\lambda/D)^2$, which corresponds to an angular aperture $2\alpha
\simeq \lambda/D$. Conversely, the solid angle under which the
source is seen \emph{from} the pinhole plane is $\Delta\Omega' =
D/z^2$, so the coherence area can also be conveniently expressed
as $A_c \simeq \lambda^2/\Delta\Omega'$. For example, the
coherence area at a distance of $1\unit{m}$ of a thermal source of
diameter $D=1\unit{mm}$ emitting at $\lambda = 0.5\unit{\mu m}$ is
$A_c\simeq 0.25\unit{mm^2}$, whereas at the same wavelength the
coherence area for the sun, which has an apparent angular diameter
$2\alpha \simeq 32'$ ($\Delta\Omega' \simeq 7\times
10^{-5}\unit{sr}$), is $A_c\simeq 4\times 10^{-3}\unit{mm^2}$.
Note that for a star like Betelgeuse ($\alpha$ Orionis), with
$2\alpha \simeq 0.047''$, $A_c$ is conversely as large as about
$6\unit{m^2}$. This last example, showing that the coherence area
of the light emitted by a star is fully coherent over the size of
our eye pupil, actually explains why stars ``twinkle'', while a
planet with a sizeable angular size does not. Of course, air
turbulence, which is the physical mechanism generating intensity
fluctuations, affects the light coming from a planet too, but
these fluctuations gets averaged out if the number of coherence
areas on our eye pupil is large.

The former considerations can be made quantitative by introducing
the key concept of \emph{mutual intensity}. Still considering a
quasi-monochromatic source, so that all delays in propagation are
much shorter than $\tau_c$, we call mutual intensity the spatial
correlation of the field at two different points
\begin{equation}\label{mutual_int}
    J_{12} = J(\textbf{r}_1,\mathbf{r}_2)= \avg{u^*(\mathbf{r}_1,t)u(\mathbf{r}_2,t)} =
    \avg{U^*(\mathbf{r}_1,t)U(\mathbf{r}_2,t)},
\end{equation}
which, when $\textbf{r}_1 = \textbf{r}_2= \textbf{r}$, becomes
just the intensity $I(\mathbf{r})$ in $\mathbf{r}$. The normalized
mutual intensity is called degree of spatial coherence
\begin{equation}\label{degr_spat_coh}
    \mu = \frac{J(\textbf{r}_1,\textbf{r}_2)}{\sqrt{I_1I_2}}.
\end{equation}
An extremely interesting result about spatial coherence comes from
considering how $J_{12}$ \emph{propagates} from a given surface,
where it is known, to another surface. The general problem is
rather complicated, but it considerably simplifies if the first
surface is actually a planar source $\Sigma$ that can be
considered as \emph{fully spatially incoherent}, by which we mean
that, over $\Sigma$,
$$ J(\boldsymbol{\rho}_1, \boldsymbol{\rho}_2) =
I(\boldsymbol{\rho}_1)\delta(\boldsymbol{\rho}_2-\boldsymbol{\rho}_1).$$
Denoting by $\boldsymbol{\rho}$ the coordinates on the source
plane, and $\mathbf{r}$ those on an observation plane further down
the propagation axis, one indeed obtains in the paraxial
approximation\footnote{Namely, for small propagation angles with
respect to the optical axis, which is the condition required for
the Fresnel approximation in diffraction to hold.}
\begin{equation}\label{VCZ}
    J(\mathbf{r}_1, \mathbf{r}_2) = \frac{\E^{-\I\psi}}{(\lambda
    z)^2}\int_\Sigma\D^2\rho\,I_0(\boldsymbol{\rho})\exp\left(\I\frac{2\pi}{\lambda
    z}\boldsymbol{\rho}\cdot \mathbf{r}\right)
\end{equation}
where $\Delta \mathbf{r} = \mathbf{r}_1-\mathbf{r}_2$ and $\psi
=\pi[r_1^2-r_2^2)]/\lambda z$. Hence, apart from a scaling and
phase factor, \emph{the mutual intensity is the Fourier transform
of the intensity distribution across the source}. Eq.~(\ref{VCZ})
is the \emph{Van Cittert-Zernike }(VCZ) \emph{theorem}, arguably
the most important result in statistical optics.\footnote{As a
matter of fact, no real source can truly be $\delta$-correlated in
space. The minimum ``physical size'' of a source is indeed of the
order of the wavelength $\lambda$, for smaller sources would emit
only \emph{evanescent} waves, exponentially decaying with the
distance from the source: hence, spatial correlations must extend
over a distance comparable to $\lambda$. Nevertheless, in terms of
\emph{propagating} waves, a source of size $\lambda$ is equivalent
to a point source.} By means of the VCZ theorem, it can be shown
that the coherence area is quantitatively given by
\begin{equation}\label{A_coh2}
    A_c = (\lambda z)^2\frac{\int|I(x,y)|^2\D x \D y}{\left|\int I(x,y)\D x \D y\right|^2}
    = \frac{(\lambda z)^2}{A_s}\frac{\avg{I^2}}{\avg{I}^2},
\end{equation}
where $A_s$ is the area of the source. For a incoherent source
with \emph{uniform} intensity (which may be an incoherently and
uniformly illuminated sample), so that $\avg{I^2} =\avg{I}^2$,
$A_c = (\lambda z)^2 /A_s$, consistently with our qualitative
approach.

The coherence area basically yields the size of the speckles
produced by a source or a diffuser around each point $P$ on the
screen. Since the field in $P$ is a random sum of the
contributions coming from all points on the source, which are
independent emitters, the total amplitude has a Gaussian
statistics. The distribution of the speckle intensity (namely, the
distribution of the intensity at different points on the screen)
is hence exponential, so there are many more ``dark'' speckles
than ``bright'' speckles. What is more important, according to the
VCZ theorem the ``granularity'' of the speckle pattern should
depend only on the \emph{geometry} of the source, and not on its
physical nature. We shall later see that this is not always
necessarily true.
\subsection{Intensity correlation}
In section~\ref{s.temcoh} we have investigated the temporal
coherence properties of optical fields. Scattering techniques,
however, usually probe \emph{intensity} correlations, which are
described by means of the normalized time--correlation function
\begin{equation}\label{g2}
    g_2(\tau) = \frac{\avg{I(t)I(t+\tau)}_t}{\avg{I(t)}^2_t}
    =\frac{\avg{u^*(t)u^*(t+\tau)u(t+\tau)u(t)}_t}{\avg{u^*(t)u(t)}^2_t}.
\end{equation}
Note that, for $\tau\rightarrow \infty$, $g_2(\tau)\rightarrow 1$,
whereas $g_1(\tau)\rightarrow 0$. While for an ideal monochromatic
source $g_2(\tau) = 1$ for all values of $\tau$, for a random
source, we should evaluate the rather complicated double sum
\begin{equation}\label{int_corr}
    \avg{I(t)I(t+\tau)} = \sum_{i,j =1}^N
\avg{u_i^*(t)u_j^*(t+\tau)u_i(t+\tau)u_j(t)}.
\end{equation}
Due to the independence of the emitters, however, a given term
averages to zero unless it contains only products of a field times
its complex conjugate relative to the \emph{same} emitter. For a
very large number $N$ of emitters, splitting the averages and
taking into account that all emitters are identical, the dominant
contribution to the sum, which is of order $N^2$, is  found  to be
$$\avg{I(t)I(t+\tau)} \simeq N^2\left[\avg{u_i^*(t)u_i(t)}^2+\left|\avg{u_i^*(t)u_i(t+\tau)}\right|^2\right],$$
which, noticing that $ N^2 \avg{u_i^*(t)u_i(t)}^2 = \avg{I(t)}^2$,
yields the important \emph{Siegert relation}:
\begin{equation}\label{Siegert}
   g_2(\tau) = 1+ |g_1(\tau)|^2.
\end{equation}
Hence, for a random source, $g_2(\tau)$ does not yield \emph{any}
additional information, and can be directly obtained from
$g_1(\tau)$; in particular, for a collision-broadened thermal
source $g_2(\tau) =1+ \exp(-2|\tau|/\tau_c)$. Nevertheless, the
distinctive difference in the long-time asymptotic behavior
between $g_2(\tau)$ and $g_1(\tau)$ yields, as we shall see, a
crucial advantage for intensity correlation techniques.

\section{Dynamic Light Scattering (Intensity Correlation Spectroscopy)}
The most popular optical correlation technique in colloid science
is  Dynamic Light Scattering, which I shall also call ``Intensity
Correlation Spectroscopy'', a denomination that captures much
better, as we shall see, the essence of the method. This short
presentation is mostly meant to stress those fundamentals of the
technique that are essential to grasp the more recent advancement
we shall later discuss. For the same reason, we shall just discuss
DLS from a system of non-interacting particles, referring to
excellent books and reviews\cite{Berne,Chu,Han,Pusey,Pusey2} for a
more comprehensive treatment.

To spot the key feature of an intensity correlation measurement,
let us make a comparison with a simple spectroscopic or
interferometric experiment, where the signal is related to the
spectrum $E(\omega)$, and therefore to the \emph{field} time
correlation function of the source, which in our case is the
scattering volume. To select a given frequency, we have to insert
a filter (such as a monochromator) on the optical path, and
\emph{then} detect the signal at the selected frequency. The basic
strategy of DLS is simply moving the filter \emph{after} the
detector, so that the photocurrent output $i(t)$ of the detector,
instead of the optical signal, is filtered. Any optical detectors
is necessarily \emph{quadratic}, namely, it detects a signal
proportional to the time--averaged \emph{intensity}
$I(t)=\overline{E^*(t)E(t)}$: hence, by using a filter whose
central frequency can be swept through a given range, the
\emph{power spectrum} of the signal can be obtained. Because of
Wiener--Kintchine theorem, an equivalent procedure is measuring
the time correlation function of $i(t)$, which is directly related
to $\avg{I(t)I(t+\tau)}$. Whatever the choice, we shall see that
operating on the photocurrent is a winning strategy for a basic
reason: at variance with field correlation spectroscopy or
interferometry the spectral bandwidth $\Delta \omega_s$ (or the
correlation time $\tau_s$) of the source illuminating the
scattering volume poses \emph{no limitation} to the measurements,
even when the spectral bandwidth of the scattered field
\mbox{$\Delta\omega \ll \Delta\omega_s$} (corresponding to a
correlation time $\tau\gg \tau_s$). The first approach, based on
using a spectrum analyzer, was mostly used at the dawn of DLS. The
invention of the digital correlator (once a complex dedicated
instrument, now just a PC data acquisition board), which allows to
work in the time domain, has however been crucial to make DLS the
spectroscopic method with the highest resolving power ever
devised.

\subsection{Time-dynamics of the scattered field}
\label{s.gaussianscatt} In a scattering experiment, the linear
dimension of the scattering volume $V$ is usually much larger than
the range $\xi$ of the structural and hydrodynamic correlations of
the systems, even when the latter extend over large spatial scales
compared to the particle size. Hence, $V$ can ideally be split
into volume elements $\delta V$ satisfying $\xi^3\ll \delta V\ll
V$. Consequently, $V$ can be regarded as a random source, where
these uncorrelated volumes $\delta V$ play the role of
``elementary emitters''. We may then expect the scattered field
and intensity to display, respectively, a gaussian and an
exponential statistics, and the time correlation functions of
$E_s$ and $I_s$ to be dictated by the temporal correlation of the
field emitted by a single elementary emitter, which will be
related to the particle dynamics in $\delta V$. There are however
a couple of warnings. First, the total scattered field has a
gaussian statistics only provided that the field scattered by each
single emitter is fully fluctuating in phase and/or amplitude.
However, this is not true for many systems of interest in colloid
science, such as glasses and gels: we shall comment on these
``nonergodic'' systems shortly. Second, the Siegert relation
connecting field and intensity correlations is violated when the
number $N$ of particles in $V$ is very small, which may be the
case when performing measurements on very diluted suspensions
under a microscope, if the coherence area of the illuminating
source is small.  In this case, by retaining the terms of order
$1/N$ in Eq.~(\ref{int_corr}), one can show that
Eq.~(\ref{Siegert}) contains an additional a \emph{number
fluctuation} term:
    \begin{equation}\label{numberfluct}
    g_2(\tau) = 1 + |g_1(\tau)|^2 + \frac{\avg{\delta N(0)\delta N(\tau)}
    }{\avg{N}^2},
\end{equation}
where $\delta N(\tau) = N(\tau) - \avg{N}$ decays on a time scale
comparable to the time it takes for a particle to move across the
scattering volume.

The field scattered by a particle suspension can be written as
\begin{equation}\label{scatt_field}
    E_s (\mathbf{q},t) = E_0 \sum_i b_i(\mathbf{q},t)\E^{\I \mathbf{q}\cdot
    \mathbf{r}_i(t)}.
\end{equation}
If particles are all identical, and provided that the scattering
amplitudes do not depend on time (which holds true for optically
isotropic particles), the normalized field correlation function is
then given by
\begin{equation*}
   g_1(q,\tau)=\frac{\avg{E^*_s (\mathbf{q},0)E_s (\mathbf{q},\tau)}}{|E_s(0)|^2}
   =F(q,\tau)\E^{-\I\omega\tau}
\end{equation*}
where we have defined the \emph{intermediate scattering function}
(ISF)
\begin{equation}\label{int_scatt_funct}
    F(q,\tau) =\avg{\sum_{i,j}
   \E^{-\I\mathbf{q}\cdot[\mathbf{r}_i(0)-\mathbf{r}_j(\tau)]}},
\end{equation}
which is nothing but the FT (in frequency) of the dynamic
structure factor $S(q,\omega)$ measured in quasi-elastic neutron
scattering experiments.\footnote{If the system is spatially
isotropic, $F(q, \tau)$ does not depend on the direction of
$\mathbf{q}$, but only on its modulus $q=|\mathbf{q}|$. In
Eq.~(\ref{int_scatt_funct}) the average is of course made over the
statistical distribution of the particle positions.} Neglecting
interactions amounts of course to assume that the position of
different particles are uncorrelated, so $g_1(q,\tau)$ is
proportional to the \emph{self} ISF
\begin{equation}\label{self intermediate scattfun}
    F_s(q,\tau) =\avg{\exp[\I\mathbf{q}\cdot\Delta \mathbf{r}(\tau)]}
\end{equation}
where $\Delta \mathbf{r}(\tau) = \mathbf{r}(\tau)-\mathbf{r}(0)$.
Therefore, $F_s(q,\tau)$ is the average value of
\mbox{$\exp[\I\mathbf{q}\cdot \Delta \mathbf{r}(\tau)]$} over the
probability distribution $p(\Delta r,\tau)$ of the particle
displacement in a time $\tau$. Note that, as a matter of fact,
$\mathbf{q}\cdot \Delta \mathbf{r}$ is just the component $\Delta
r_q$ of the particle displacement \emph{in the direction of the
wave-vector} $\mathbf{q}$. Hence, $F_s(q,\tau)$ can be seen as the
Fourier transform $\mathscr{F}[p(\Delta r_q, \tau)]$, which is the
\emph{characteristic function} of $p(\Delta r_q,\tau)$.  Given the
characteristic function, all the moments of a probability
distribution are easily calculated. For instance, the mean square
particle displacement along $\mathbf{q}$ is given by
\begin{equation}\label{ms-displ}
    \avg{\Delta^2 r_q(\tau)} = -\left[\frac{\partial g_1(q,\tau) }{\partial
    q^2}\right]_{q=0}
\end{equation}
\subsection{Time-correlation of the field scattered by Brownian
particles} The simplest model of a freely--diffusing Brownian
particle is that of a mathematical random walk. In one dimension,
the particle motion is seen as a sequences of random ``steps''
$x_i$ along the positive or negative direction, so that $\avg{x_i}
=0$ and, if we assume the steps to be uncorrelated $\avg{x_ix_j} =
\avg{x_i^2}\delta_{ij} = \Delta^2 \delta_{ij}$. Then, because of
the Central Limit Theorem, the total displacement $x= \sum_{i=1}^N
x_i$ for a large number $N$ of steps is a gaussian random variable
with $\avg{x}=0$ and $\sigma_x^2 = \avg{x^2} = N\Delta^2$. This
corresponds, in a continuum description, to a diffusion process
with a diffusion coefficient $D= \Delta^2/2\Delta t$, where
$\Delta t$ is the time it takes for a step. Generalizing to 3D,
the particle mean square displacement is then given by
$\avg{\mathbf{r}^2(t)} = 6Dt$, where, because of the celebrated
Einstein's relation, the diffusion coefficient is related  to the
hydrodynamic friction coefficient\footnote{$\zeta = 6\pi\eta a$
for a spherical particle of radius $a$ in a solvent of viscosity
$\eta$.} $\zeta$ by $D=k_BT/\zeta$ .

For $t\rightarrow 0$, the random walk model yields however a
rather unphysical result, because the particle velocity diverges
as $t^{-1/2}$. A more consistent description is obtained from the
Langevin equation,\cite{Kubo} whose solution shows that the
particle motion becomes diffusive only after the
\emph{hydrodynamic relaxation time} $\tau_B = m/\zeta$, where $m$
is the particle mass, which is the decay time of the velocity
time-correlation function. It is also useful to note that the
diffusion coefficient is just the time integral of the latter
\begin{equation}\label{velcorr}
   D = \frac{1}{3}\int_0^\infty \!\!\!\avg{\mathbf{v}(0)\cdot\mathbf{v}(t)}\D t
\end{equation}
For $t\gg \tau_B$, the probability for a particle to be in
$\mathbf{r}$ if it was in the origin at \mbox{$t=0$} is then a
gaussian. Note however that we need only the component of the
displacement in direction of $\mathbf{q}$ (which can in fact be
taken as the $x$ axis), hence $p(\Delta r_q, \tau)$ is a gaussian
with $\avg{\Delta  r_q} =0$ and variance $\sigma^2 = 2D\tau$.
Being the characteristic function of a gaussian centered on the
origin, $F_s(q,t)$ is itself a gaussian in $q$ with variance
$1/\sigma^2 = (2D\tau)^{-1}$, $ F_s(q,\tau) = \exp(-Dtq^2)$. Then
as a function of $\tau$, the ISF decays exponentially with a rate
$\Gamma = Dq^2$. The field and (because of the Siegert relation)
the intensity correlation functions are given by
\begin{equation}\label{SINGLE_PART}
    \left\{\begin{array}{l}
             g_1(\tau) = \exp(-\I\omega t)\exp(-\Gamma \tau) \\
             g_2(\tau)= 1+ \exp(-2\Gamma \tau).\\
           \end{array}\right.
\end{equation}

\subsection{DLS, the ultimate spectroscopy}
Brownian motion gives then rise to a spectral broadening $\Gamma =
Dq^2$ that, because $D$ is related to the particle radius, should
allow for particle sizing. The problem, however, is that these
spectral broadenings are extremely small, because colloidal
diffusion is extremely slow:  for instance, expressing its radius
$R$ in nanometers, a spherical particle in water at $20^\circ$C
has $D\simeq (2.15/R)\times 10^{-6} \unit{cm^2/s}$. Since the
largest accessible $q$-values in light scattering are about
$3\times 10^5\unit{cm^{-1}}$, even for a small surfactant micelle
with a radius $R = 2\unit{nm}$ the spectral broadening is of the
order of $0.1 \unit{MHz}$, which is negligible compared to the
bandwidth of a spectral lamp, or of a common laser with no
longitudinal mode selection. For ``usual'' colloids with a size in
the tenths of a micron range, the situation is obviously far
worse. Measuring a spectral broadening that is much smaller than
the  source intrinsic bandwidth is of course extremely
challenging: as a matter of fact, it is totally out of question
for any spectroscopic method relying on \emph{field} correlations.

Yet, things change dramatically if we consider \emph{intensity}
correlations. This is probability easier to see in the time
domain. Assume that a source has a bandwidth $\Delta \omega_s$,
hence a coherence length $\ell_c \simeq 2\pi c/\Delta\omega$. If
the scattering volume has linear dimensions $\ell =(V_s)^{1/3} \ll
\ell_c$, which is usually the case,\footnote{Even for a bandwidth
of the order of the GHz, $\ell_c$ is of the order of a few
centimeters.} each point in $V_s$ basically ``sees'' the same
incident field. Hence, we can write $\mathbf{E}_s(q,t) = B(q,t)
\mathbf{E}_0(t)$, where $\mathbf{E}_0(t)$ is the incident field
and $B(q,t) = \sum_i b_i(q)\exp[\I \mathbf{q}\cdot
\mathbf{r}_i(t)]$ the total scattering amplitude. However,
$\mathbf{E}_0(t)$ and $B(q,t)$ are clearly independent random
variables, so we have: $\avg{B^*(q,0)E^*_0(0) B(q,\tau)E_0(\tau)}
= \avg{B^*(q,0)B(q,\tau)}\avg{E^*_0(0)E_0(\tau)}.$ Hence, the
field
 correlation function \emph{factorizes} as
\begin{equation*}
    g_1(q,\tau) = g_1^S(\tau)g_1^B(q,\tau)
\end{equation*}
where $g_1^S(\tau)$  is the time correlation function of the
source and $g_1^B(q,\tau)$ is the sample correlation function due
to particle Brownian motion. Since $g_1^S(\tau)$ decays to zero on
the correlation time $\tau_c$ of the source, which is far shorter
than the Brownian correlation time, there is no way to follow the
decay of $g_1^B$. Consider however the \emph{intensity}
correlation function. Again, we can write
\begin{equation*}
    g_2(q,\tau) = g_2^S(\tau)g_2^B(q,\tau)
\end{equation*}
Yet, in this case, for $\tau\gg \tau_c$, $g_2^S(\tau)$ decays to
\emph{one}, and we have:
\begin{equation}
    g_2(q,\tau) \underset{t\gg \tau_c}{\longrightarrow} g_2^B(q,\tau)
\end{equation}
which is \emph{exactly} what we want to measure.
\begin{figure}[h]
\centering
  \includegraphics[width=\columnwidth]{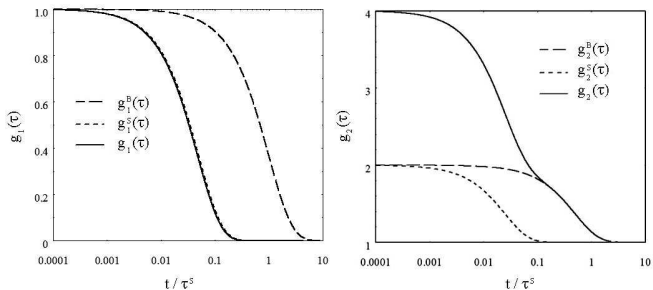}
  \caption{\label{f1} Behavior of the field (left) and intensity (right) correlation functions, using a
  temporally partial coherent thermal source with $\tau_c^S = 0.05 \tau_c^B$.}
\end{figure}
In other words, we actually want to \emph{avoid} using a source
with a very long coherence time, for we need $\tau_c$ to be much
shorter than the physical fluctuation time of the
sample.\footnote{Note that $g_2(\tau)$ decreases from \emph{four}
to one because the scattered field is, at least in the case of a
pure thermal source, the product of \emph{two} gaussian
processes.}

Of course, using single longitudinal mode lasers $g_2^S \equiv 1$,
even if the effective laser bandwidth is not negligible, because
the spectral broadening is due to pure phase fluctuations. The
latter, however, \emph{still} affect $g_1(\tau)$, thus hampering
spectroscopic and interferometric measurements.
Quantitatively,\cite{Pusey} one finds that the scattered field is
\emph{not} gaussian, so that, in terms of the full correlation
functions $g_2(q,\tau)\ne 1+|g_1(q,t)|^2$; yet, $g_2(\tau)=
1+|g_1^B(q,t)|^2$, thus intensity correlation measurements still
yield what is needed. Even if useful, using single--mode lasers in
DLS is not at all compulsory,  so much that the first attempts to
study Brownian motion by analyzing the intensity fluctuations of
speckle patterns were performed by Raman using a conventional
mercury-arc lamp.\cite{Raman} Hence, lasers are not used in DLS
setups because of they are particularly monochromatic but, as we
shall shortly see, just for practical reasons related to their
unique \emph{spatial} coherence properties.

In the frequency domain, we can see that the ``magic'' of
intensity correlation comes from the fact that doing DLS is like
playing a kind of  ``optical radio''. To broadcast an audio signal
$v_s =f (t)$ we can for instance modulate the amplitude of a
carrier wave at a radio frequency $\omega_c$ much larger than the
frequency components of $f(t)$:
$$v(t) =A[1+mf(t)]\cos \omega_c t.$$
Then,  to ``decode'' the signal, we use again a \emph{quadratic}
detector, which basically consist of a rectifier (a simple galena
crystal in the first radios, a diode later). Suppose for
simplicity that we wish to transmit a simple sinusoidal signal
$\cos \omega_m t$, with $\omega_m \ll \omega_c$. Before the
rectifier, the broadcast field is:
$$ v(t) = A\cos\omega_c t+ \frac{mA}{2} [\cos(\omega_c+\omega_m)t
+\cos(\omega_c-\omega_m)t].$$ This contains, besides the original
carrier frequency, two symmetric sidebands with \mbox{$\Delta
\omega =\pm \omega_m$} but,  because $\Delta \omega \ll \omega_c$,
no resonant filter can resolve them. After the rectifier,
supposing that the modulation depth $m$ is small, we have:
$$ v^2(t) \simeq
\frac{A^2}{2}[1+\cos 2\omega_ct + m\cos(2\omega_c\pm\omega_m)t] +
mA^2 \cos \omega_mt,$$ namely, besides a zero--frequency component
and three components at radio-frequency (RF), we have obtained a
signal at the \emph{modulation} frequency that can be extracted
with a low-pass filter. This strategy, which is called homodyne
detection (the signal is ``mixed with itself''), is again the
result of using a quadratic detector. In DLS, the photodetector
plays a role quite similar to the galena crystal, with $B(q,t)$ as
modulating signal, although in the form $f(t)v_c(t)$ instead of
[$1+mf(t)]v_c(t)$.\footnote{The exact analogous in radio
engineering is dubbed ``carrier--suppressed AM''.} The net effect
of the ``self--beating'' of the scattered field on the quadratic
detector is reconstructing a copy of the spectrum of $B(q,t)$
\emph{in baseband}, but with all frequencies doubled.

\subsection{Spatial coherence requirements in DLS}
Intensity correlation measurements have several requirements in
terms of spatial coherence for what concerns both the illuminating
source and the detection scheme. Maximizing the DLS signal
requires indeed to illuminate the scattering volume with a
spatially coherent beam. Yet, we have seen that a source of area
$A$ emits a spatially coherent field only within a solid angle
$\Delta \Omega \simeq \lambda^2/A$: the useful emitted power is
then just the amount contained in $\Delta \Omega$, namely, $P
=SL\Delta \Omega$, where $L$, the power emitted per unit area and
solid angle, is the \emph{radiance} of the source (sometimes also
called ``brightness'', or ``brilliance''). The crucial difference
between a laser and a spectral lamp is actually its enormously
higher \emph{spatial} coherence, which is strictly related to its
directionality. In fact, a gaussian beam emitted by a laser is
perfectly coherent over its whole section, and diverges with the
diffraction angle $\Delta \Omega = \lambda^2/w_0^2$, where $w_0$
is the minimum beam--spot size. The section of the emitted beam
can therefore be regarded as a ``speckle'' emitted by a source of
size $w_0$; a source, however, that emits all its power on a
\emph{single} speckle. It is actually their \emph{high brilliance}
that make lasers practically indispensable in DLS.

Let us now consider  detection. The scattering volume behaves as a
random source, with a size that is just the projection
perpendicular to $\mathbf{q}$ of the illuminated volume. As a
consequence, there is no advantage in using a detector with an
area $A$ larger than a coherence area of this source. Namely,
increasing the detector area beyond the size of the speckles made
by the scattered field  increases the detected power, but this
additional power is of no use, for different speckles are
uncorrelated. If the number $N =A/A_c$  of collected speckles is
large, intensity fluctuations will grow just as $ \sigma(I) \sim
N^{1/2}$ (it is a Poisson statistics). Hence $g_2(0) -g_2(\infty)
= \sigma^2(I)/\avg{I}^2\sim N^{-1},$ so we just loose contrast.
For a generic value of $N$, one can actually write a ``corrected''
Siegert relation of the kind $g_2(\tau) = 1+f(N)|g_1(\tau)|^2$,
where the spatial coherence factor $f(N)$ can be approximately
written as $f(N) \simeq (1+N)^{-1}$. To get a high contrast (a
``good intercept'', in the jargon of DLS) , the detector aperture
should be \emph{considerably} smaller then a coherence area.

In the earliest schemes of a DLS apparatus, the angular extent of
the scattered light reaching the photodetector was limited by
means of two pinholes aligned along the selected scattering
direction. However, a much more efficient detection scheme, which
consists in forming by a lens an image of the scattering volume on
a slit that can be closed or opened by micrometers to select a
single speckle, was soon adopted. The real novelty is that the
effective size of speckle on the slits can be tuned by
stopping--down the lens with an iris diaphragm, because the image
of a speckle gets convoluted with the lens pupil, so that by
reducing the lens aperture the size of a coherence area on the
image plane increases.\cite{GoodmanFO} We shall return to this
idea of performing a ``spatial coarse-graining'' on the image
plane in section~\ref{s.imaging}. With these ``traditional''
detection schemes it is however very hard to reach a condition
close to the ``ideal'' contrast $g_2(0) -g_2(\infty) =1$, which is
conversely ensured by novel detection schemes using single-mode
fibers that have become widespread in the last two decades.
Understanding fiber detection requires however to forget all about
``geometrical'' arguments: neither the size of the fiber to be
used, nor the distance of its opening from the sample, have indeed
anything to do with the speckle size. Rather, an optical fiber has
to be regarded as  an ``antenna'', which can resonate only on
well-defined proper ``modes''. A \emph{monomode} fiber, in
particular, allows for a single propagating mode, whose spatial
structure is very similar to the fundamental transversal mode of a
laser and display therefore full spatial coherence. The field
detected by such a fiber is nothing but the \emph{projection} (in
the full mathematical sense) of the scattered field on the single
fiber mode. The amplitude of the field collected by the fiber can
vary by changing the size of the scattering volume or of the fiber
core but, because of the full spatial coherence of the fiber mode,
the field and intensity correlation functions always show
\emph{full contrast}, with values $g_1(0) =1$ and $g_2(0)=2$ at
zero delay. One can show that the amplitude of the projected
component can be maximized by matching the angular aperture of a
speckle with the acceptance angle of the fiber. Besides being much
simpler both conceptually and practically,  fibers receivers
present another very interested feature: if a laser beam is fed
into the fiber from the \emph{opposite} terminal (the one usually
bringing the collected light to the photodetector) and launched
towards the scattering cell from the receiver input, its spatial
intersection with the incident beam allows to precisely define the
scattering volume. By this trick, optical alignment, which is
time--consuming in traditional DLS setups, becomes much
simpler.\cite{Ricka}

\subsection{Heterodyne detection and Doppler velocimetry}
In radio engineering, homodyne detection has the disadvantage of
generating a signal at $\omega_m$ which is proportional to the
(generally weak) amplitude of the carrier wave detected by an
aerial. Radios became much more efficient with the development of
the ``heterodyne'' receiver, where the signal power is ``pumped
up'' by mixing it with the signal $v_L(t) = A_L\cos\omega_ct$ from
a \emph{local oscillator} (LO) at the frequency of  the carrier
wave. Indeed, using a mixer that multiplies the incoming and LO
signals, we get again the audio signal, but \emph{amplified} by
$v_L$:
$$ V(t) = v(t)v_L(t) = AA_L(1+\cos \omega_mt)\cos^2\omega_p(t) =
\{\mathrm{RF signals}\} + AA_L\cos\omega_mt.$$ A very similar
trick is used in heterodyne DLS, where the LO is simply a fraction
of the incident beam (even simply a reflection from the cell
windows) which ``beats'' with the scattered field on the
photodetector. We have then
$$\avg{I(0)I(\tau)}^{HD} = \avg{|E_s(0)+E_L(0)|^2
|E_s(\tau)+E_L(\tau)|^2}.$$ Neglecting fluctuations in the
incident field (hence in $E_L$), observing that $E_L$ and $E_s$
are uncorrelated, and  assuming that $|E_L| \gg |E_s|$ (which is
almost unavoidable), one obtains after some calculation
\begin{equation}\label{heterodyne}
    g_2^{HD}(\tau) = 1 + k\mathrm{Re}[g_1(\tau)]
\end{equation}
where $k = \avg{I_s}/I_L$.  The important difference with respect
to homodyne DLS is that, by heterodyning, we also detect the real
part of oscillating terms of the form  $\exp(\I\omega \tau)$.
Consider for instance a colloidal suspension in flow with a
uniform velocity $\mathbf{v}$. The field correlation function can
be evaluated by adding to the diffusion equation an advective term
$\mathbf{v}\cdot \boldsymbol{\nabla} c$. Using the same method we
have described earlier, one finds $ g_1(q,\tau) = \exp(\I
\mathbf{q}\cdot \mathbf{v} \tau)\exp(-Dq^2\tau).$ The first phase
term is totally ``invisible'' in homodyne detection, whereas:
$$ g_2^{HD}(\tau) = 1 + k\exp(-Dq^2\tau)\cos (\mathbf{q}\cdot
\mathbf{v}\tau)$$ Heterodyne detection is therefore at the roots
of \emph{Laser Doppler Velocimetry}, which allows to study
hydrodynamic motion using particles as tracers, or the drift
particle motion induced by an external field, such as in
electrophoresis.

\section{Novel investigation methods based on
intensity correlation}

\subsection{Multi--speckle DLS and Time-Resolved Correlation (TRC)}
Colloidal gels and glasses are a class of materials of prominent
interest characterized by an extremely low, quasi--arrested
dynamics where each single particle performs a restricted motion
around a fixed position. Because of the limited particle
displacement, the scattered field can be written as the sum
$\mathbf{E}_s(q,t) = \mathbf{E}_f(q,t) + \mathbf{E}_c(q)$ of a
fully fluctuating component $\mathbf{E}_f(q,t)$ plus a
time--independent contribution $\mathbf{E}_c(q)$. As a main
consequence, $\mathbf{E}_s(q,t)$ is not anymore a
fully--fluctuating gaussian random variable, and its statistical
properties of are very different from those of the light scattered
by free Brownian particles. The value of $\mathbf{E}_c(q)$ depends
indeed on the specific configuration of the scatterers as seen
from a given detection point, hence it is different from speckle
to speckle because each coherence area comes from a unique
combination of the phases of the individual fields scattered by
each particle. Therefore, while evaluating the ensemble average of
the scattered field over many speckles we get
$\avg{\mathbf{E}_s(q,t)}_e = 0$, the \emph{time} average of
$E_s(q,t)$ does not vanish, but is rather given by
\mbox{$\avg{\mathbf{E}_s(q,t)}_t = \mathbf{E}_c(q)$}. Retrieving
sound structural information by DLS  on gels and glasses requires
then to measure ensemble--averaged correlation functions. The
latter can be of course obtained with a ``brute force'' method by
very slowly displacing or rotating the cell between distinct
acquisitions of $g_2(t)$, so that the detector is sequentially
illuminated by many independent speckles. A different and far less
time--consuming strategy  was however proposed by Pusey and van
Megen, who showed that the correct, ensemble-averaged correlation
function may be reconstructed from the intensity correlation
function measured in a \emph{single} run on a \emph{fixed}
speckle, provided that  the ensemble--average of just the
\emph{static} intensity $\avg{I}_E$ is carefully measured. The
correct intensity correlation function is obtained from  the
single-run $g_2(\tau)$ and the ratio $\avg{I}_t/\avg{I}_E$ with a
well--defined, although non trivial, correction scheme.\cite{PvM}

Investigating ``non-ergodic'' media by traditional DLS is anyway
laborious.  Luckily, we can actually take advantage from the very
slow dynamics of colloidal gels and glasses. In fact, neither a
fast detectors as a photomultiplier, nor a real--time digital
correlator are needed:  a digital camera  with a moderately fast
data acquisition and transfer rate fully suffices, and the
calculation of $g_2$ can still be made in real time via software.
CCD and CMOS cameras are moreover multi-pixel devices, where each
pixel acts as a detector, hence, in principle, we have a way to
perform DLS measurements simultaneously on a vary large number of
speckles. The outcome of such a multi-speckle experiment is a
series of speckle images, where the intensity for each pixel $p$
and time $t$ is recorded. The intensity correlation function is
then obtained as
$$ g_2(\tau) =
\avg{\frac{\avg{I_p(t)I_p(t+\tau)}_p}{\avg{I_p(t)}_p\avg{I_p(t+\tau)}_p}}_t$$
where  $\avg{\cdots}_t$ is a time average, whereas
$\avg{\cdots}_p$ denotes an average over an appropriate set of
pixels corresponding to the same $q$-value.\footnote{The order in
which these two averages is taken is crucial to obtain a correct
ensemble--averaged $g_2(\tau)$. This is evident for
fully--arrested sample, where one expects $g_2(\tau) \equiv 1$ for
all $\tau$, whereas $I$ is constant in time but varies from pixel
to pixel, so that reversing the order of averaging we would obtain
$g_2(\tau)\equiv 1$.} Because of the pre--averaging over many
pixels, yielding very smooth data, multi--speckle detection yields
a tremendous reduction of measurement time.

Multi--speckle methods are also ideal for investigating systems
displaying \emph{heterogeneous} temporal dynamics in glasses,
foams, and a variety of jammed systems that often evolve in time
through intermittent rearrangements. This is the principle of the
Time--Resolved Correlation (TRC)
technique,\cite{Cipelletti1,Cipelletti2} where the change of the
sample configuration is obtained by calculating the degree of
intensity correlation between pairs of images taken at time $t$
and $t + \tau$, which explicitly depends on $t$
$$ c_I(t,\tau) =
\frac{\avg{I_p(t)I_p(t+\tau)}_p}{\avg{I_p(t)}_p\avg{I_p(t+\tau)}_p}
-1.$$ The
 amplitude of the fluctuations in the temporal dynamics can then
be quantified by the variance $\chi(\tau)= \avg{c^2_I(t,\tau)}-
\avg{c_I(t,\tau)}^2$, which is directly related to the so-called
dynamical susceptibility $\chi_4$ used to characterize dynamic
heterogeneity in computer simulations of the glassy state. In the
last section we will see that an extension of TRC, allowing to
resolve $g_2(\tau)$ both  in time and space, provides a basic link
between scattering and imaging.

\subsection{Near Field Scattering (NFS)} \label{NearField}
\begin{figure}[h]
\centering
  \includegraphics[width=\columnwidth]{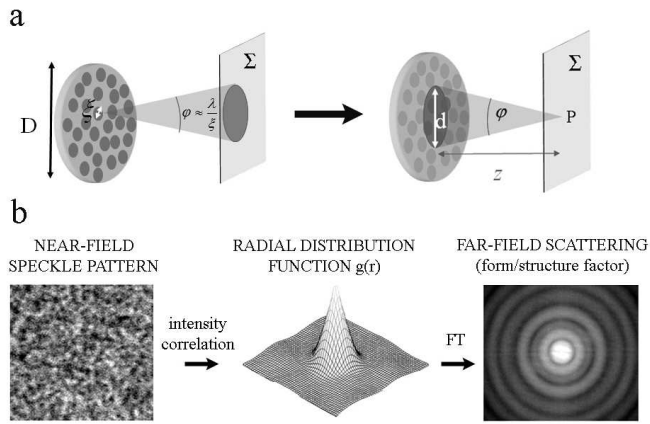}
  \caption{\label{f2} Speckles in near field (a) and sketch of a NFS experiment (b).}
\end{figure}

Because the scattered intensity has (for ergodic media) an
exponential distribution  with $\avg{I^2}= \avg{I}$,
Eq.~(\ref{A_coh2}) basically states that the size of a speckle is
just fixed by the geometry of the scattering volume, and does not
contain any information about the physical mechanisms that produce
scattering (see section~\ref{s.spacoh}). This is a consequence of
the VCZ theorem, which is however strictly valid only when the
source is not spatially correlated.  In fact, it is definitely
\emph{not} true for by a ``structured source'', by which we mean a
sample scattering light because of the presence of correlated
regions of size $\xi \gtrsim \lambda$, due for instance to an
inhomogeneous refractive index distribution.\cite{Giglio1} For
example consider, as in Fig.~\ref{f2}a, the scattering pattern
generated on a close-by plane at distance $z$ from the cell by a
suspension of colloidal particles contained in a thin cell, and
illuminated with a beam spot of diameter $D$. Particles with a
size  $\xi \gtrsim \lambda$ scatter light mostly within a cone of
angular aperture $\varphi \simeq \lambda/\xi$ (which, for very
large particles, coincides with the angular aperture of their
diffraction pattern). By reciprocity, light can reach a given
point $P$ on the observation plane \emph{only} from a region of
size $d\simeq z\varphi$. Hence, $P$ sees an ``effective'' source
with a size that, provided that $z< z_c = D\xi/\lambda$, is
\emph{smaller} than $D$. Rather surprisingly, the speckles
generated by such a source have a typical dimension
$$\frac{\lambda}{d} z = \lambda \left(\frac{\xi}{\lambda z}\right)z = \xi.$$
The statistical size of a near--field speckle (which according to
the VCZ theorem should \emph{vanish} for $z\rightarrow 0$) is
therefore of the order of the particle size.

More quantitatively, it turns out that, for a structured source
with a generic mass distribution, the intensity correlation
function of the scattered light in the near--field  is
proportional to the radial distribution function $g(r)$, which
yields, for non-interacting scatterers with a finite size, the
average value for the speckle size we found with the former
qualitative argument.\cite{Giglio2} The intensity distribution
$I(q)$ measured in the usual far--field scattering experiments,
which is conversely proportional to the structure factor of the
sample, can then also be obtained by evaluating the \emph{power
spectrum} of the intensity on a near field plane. These
conclusions are fully confirmed by a reassessment of the VCZ
theorem for a source with finite spatial correlation, which leads
to conclude that, within the so-called ``deep Fresnel region''
(DFR) \mbox{$z< z_c$}, corresponding to large Fresnel
numbers,\footnote{\label{regions}We recall that, in diffraction
optics, the far-field Fraunhofer diffraction pattern from a source
of size $D$ is observed only for $z\gg z_F =D^2/\lambda$, whereas
the more complex Fresnel diffraction regime corresponds to $z<
z_F$. Since it is easy to see that $z_c \ll z_F$, the near--field
region always lies ``deep'' within the latter (hence the name).}
the field correlation function is actually invariant upon
propagation and approximately equal to that on the source plane,
so the speckles \emph{retain the same size all along this region}.
For \mbox{$z\gg z_c$}, conversely, the source basically act as a
collection of $\delta$-correlated emitters, and the standard VCZ
theorem yields a good approximation for the mutual intensity on
the observation plane.\cite{Cerbino1}

This Near-Field Scattering (NFS) technique present several
advantages with respect to traditional methods to measure
small-angle scattering, in particular when made using a heterodyne
detection scheme, which just consists in letting the scattered
field to ``beat'' with the transmitted beam, without blocking the
latter:\cite{Giglio2} in this configuration it requires indeed an
extremely simple optical setup, in principle just a multi--pixel
detector placed on the near-field observation plane. Of course,
the speckle size should not be much smaller than the size of a
pixel of the sensor, since we would otherwise average over many
uncorrelated speckles, loosing contrast. However, if this
condition is not met, the speckles can be magnified using a
microscope objective: actually, the speckle size on the image
plane depends only on the numerical aperture NA of the objective,
and can be enlarged at will by reducing the latter. This trick of
magnifying speckles by just stopping-down the imaging optics, is
in fact similar to  what is done in DLS detections by closing the
diaphragm of the lens that images the sample volume on the slits.
A second important advantage is that, because the scattered and
transmitted beams are perfectly superimposed, NFS is an ideal
heterodyne method that provides an \emph{absolute} measure of the
scattering cross sections, since the strength of the local
oscillator is exactly known. An example of NFS experiment, made in
our lab to obtain the form factor of very diluted polystyrene
particles, in shown in Fig.~\ref{f2}b.

\subsection{NFS velocimetry} \label{s.NFSvelo} Besides providing a simple and
efficient tool to obtain the structure factor of a suspension at
very small angles, heterodyne NFS can be used as a very accurate
technique to measure the local motion in a fluid, using colloidal
particles as ``tracers'' like in  Particle Imaging Velocimetry
(PIV\cite{Adrian}). In a PIV measurement, a fluid containing
tracer particles is illuminated by a thin sheet of light and
imaged in the perpendicular direction. By measuring the tracer
displacement between two closely spaced times, the
\emph{two-dimensional} in-plane velocity of the fluid is
recovered, whereas a full 3-D reconstruction of the field profile
can be obtained by holographic methods.\cite{Adrian2} Of course,
tracking individual particles requires the latter to be large
enough to be imaged, namely, the particle size must be larger than
the resolution limit of the imaging system. This is usually
acceptable when studying macroscopic hydrodynamics flow, but may
raise several problems when dealing with flow around very small
structures, which is often the case in microfluidic experiments.
Suppose however that we perform a NFS measurement on a moving
suspension, namely from particles that, besides performing
Brownian motion, are transported by the suspending fluid. As we
discussed, a detector placed on a plane $P$ within the
deep-Fresnel region, or on the plane where $P$ is imaged by a
microscope objective, collects light from a region $D^* \ll D$,
where $D^*$ is respectively determined by the scattering cone of
the scatterers or by the NA of objective. If all scatterers are
rigidly displaced transversally to the optical axis, the speckle
field just displaced accordingly, with no relative change in the
speckle position.\footnote{Provided at least that the particles
generating a given speckle are subjected to a constant
illumination, a condition that is met provided that $D^*$ is well
inside $D$.} Note that this one-to-one mapping between particle
motions and speckles displacement works only in NFS conditions:
upon particle motions, the far--field speckle pattern remains
\emph{stationary}, simply fluctuating in time due to Brownian
motion, because each speckle is the result of contributions
arriving from the \emph{whole} illuminated region $D$.

Hence, a statistical analysis of speckle patterns taken at
different times allows to recover the tracer motion, and map the
fluid velocity profile.\cite{Alaimo} This can either be done by
measuring the cross-correlation function between two subsequent
patterns, or from observing the effects of the tracer motion on
the far--field scattered intensity reconstructed by a Fourier
transform. Writing the total heterodyne intensity as
$I(\mathbf{r},t) = I_0 + \delta I_t(\mathbf{r})$, where
\mbox{$\delta I_t(\mathbf{r}) =2\mathrm{Re}[E_tE_s^*(t)]$}, and
assuming that the fluid embedding the tracers is moving at
constant velocity $\mathbf{V}$, after a delay $\Delta t$ the
fluctuating part becomes $\delta_{t+\Delta t} (\mathbf{r})$ =
$\delta_t (\mathbf{r}-\Delta\mathbf{r})$, where
$\Delta\mathbf{r}=\mathbf{V}\delta t$. Then, the cross-correlation
of the speckle pattern between $t$ and $t+\Delta t$ is simply a
``shifted version'' of the signal at time $t$:
\begin{equation}\label{veloc_corr}
    G_{\Delta t} (\mathbf{x}) = \avg{\delta I_t(\mathbf{r})\delta I_{t+\delta
t}(\mathbf{r}+\mathbf{x})}= \avg{\delta I_t(\mathbf{r})\delta
I_{t}(\mathbf{r}-\Delta \mathbf{r} +\mathbf{x})}= G_0
(\textbf{x}-\Delta
    \mathbf{r}).
\end{equation}
In other words, the cross-correlation shows a pronounced peak
located at $\mathbf{x}= \Delta \mathbf{r}$ that, for constant
$\mathbf{V}$, shift linearly with time. For practical reasons, it
is often more useful considering the autocorrelation of the
difference signal $$\delta I'_{\Delta t} = I_{t+\Delta
t}(\mathbf{r}) - I_{t}(\mathbf{r}) = \delta I_{t+\Delta
t}(\mathbf{r}) - \delta I_{t}(\mathbf{r}),$$ which is a
zero-average fluctuating variable that does not require, to be
evaluated, the subtraction of the time--independent background. In
this case, one gets \emph{two} symmetric correlation
peaks.\cite{Alaimo} This alternative approach is also useful
because, considering the FT of $\delta I'_{\Delta t}$ and making
use of the shift theorem, one easily finds that
\begin{equation}\label{veloc_struct}
   \left|\mathscr{F}[\delta I'_{\Delta t}]\right|^2 = I(q)[1-\cos (\mathbf{q}\cdot
\Delta\mathbf{r})].
\end{equation}
Thus, particle motion shows up in the structure factor as a set of
straight fringes perpendicular to $\mathbf{V}$, with a spacing
$\Lambda = 2\pi / |\mathbf{q}\cdot \mathbf{V} \Delta t|$ that
narrows linearly in time.

\subsection{The third dimension of the speckles}
In section~\ref{s.spacoh} we have discussed the two-dimensional
properties of the speckles, namely, what is the statistical
distribution and characteristic ``granularity'' of the maculated
pattern observed on a screen placed at a given distance from a
random source. In fact, we have spoken of coherence \emph{areas}:
however, we may wonder whether speckles also have a ``depth''
along the direction of propagation, and how this depth depends on
the distance from the source. To avoid confusion, we are referring
here to a purely \emph{spatial} longitudinal coherence for a
monochromatic source (for a polychromatic source with finite
bandwidth, there is an obvious longitudinal limit to extent of
field correlations, which is given by the coherence length
$\ell_c$). The longitudinal coherence of speckles is important in
several novel techniques, such as speckle photography,
interferometry, and holography, yet it has been the subject of
relatively few theoretical investigations (for a review till 2007,
see.\cite{GoodmanSPE})

Without entering in the details of the analysis, we just quote
here the main results obtained in  novel
approaches\cite{Alaimo2,Gatti} where the problem is carefully
reconsidered in relation to the distance from the source: quite
different properties of the 3-D speckles are indeed found
depending on whether they are observed in the deep Fresnel region,
in far-field Fraunhofer diffraction, or in the intermediate
``full'' Fresnel regime where the VCZ theorem already holds in the
form given by Eq.~(\ref{VCZ})  (see footnote~\ref{regions}).
Suppose that a random diffuser is illuminated by a laser beam
\emph{focused} on the diffuser to a spot size $D$ (so that the
illuminating wavefront is flat\footnote{\label{gauss}We recall
that the spot size $w(z)$ and radius of curvature of a gaussian
laser beam focused in $z=0$ to a minimum spot size (beam waist)
$w_0$ are given by $w(z) = w_0 \sqrt{ 1+ (z/z_R)^2 }$, $R(z) =
z\left[1+(z_R/z)^2\right]$, where $z_R = \pi w_0^2/\lambda$ is
called the Rayleigh range. Hence, the wavefront at $z=0$ is flat,
whereas both $R(z)$ and $w(z)$ grow linearly with $z$ for $z\gg
z_R$, corresponding to an angular divergence of the beam $\theta
\simeq \lambda/(\pi w_0)$. Note that $w(\pm z_R) = w_0\sqrt{2}$,
so that within the Rayleigh range, the cross-section of the beam
changes only of a factor $\sqrt{2}$, whereas the curvature radius
is \emph{maximal} at the Rayleigh range, $R(\pm z_R) = 2z_R$.}).
We shall also assume that the source is \emph{quasi--homogeneous},
meaning by that that the spatial correlations of the diffuser
extend over a typical size $\xi \ll D$. Then, the transitions
between a different ``morphology'' of the 3-D speckles generated
by the diffuser are marked by the distances $z_c =D\xi/\lambda$
and $z_F = D^2/\lambda$. Let us summarize the main aspects of
these regimes.

\paragraph{``Deep'' Fresnel region ($z<z_c\ll z_F$)} As already
discussed, in this near--field region the trasverse coherence
length $\delta x$ of the speckles does not depend on $z$ and
coincides with $\xi$. Physically, the speckle pattern on a plane
at distance $z$  can be pictures as made of luminous ``spots''
with an average size $\xi$, separated by a typical distance which
is also of order $\xi$. The \emph{longitudinal} coherence length
$\delta z$ can be qualitatively found as follows.  Since $z \ll
z_F$, the beam wavefront is still approximately flat, namely, each
speckle behaves like an aperture illuminated by a plane wave and
broadens upon propagation just because of diffraction, with a
characteristic diffraction angle $\vartheta_d\sim \lambda/\xi$.
The longitudinal coherence length can be roughly evaluated as the
distance where the diffraction patterns from two neighbor speckles
starts to interfere. Hence $\delta z \simeq \xi/\vartheta_d\simeq
\xi^2/\lambda$, which correctly estimates the value obtained from
a rigorous approach. A 3-D speckle in near field can then pictured
as a kind of ``jelly bean'' with a coherence volume of order $
\xi^2\times (\xi^2/\lambda) = \xi^4/\lambda.$

\paragraph{Fraunhofer region ($z\gg z_F$)} As the distance from the
source
    increases beyond $z_c$,  we enter the region where the usual VCZ theorem
    holds.
    Here speckles grow in transverse size as $\delta x(z) \simeq \lambda z/D$, so
    they diffract at smaller and smaller angles
    $\vartheta_d \simeq \lambda/\delta x \simeq D/z$. On the other
hand, the wavefront of the overall beam becomes progressively
\emph{curved}, so that each speckle ``expands'' as a spherical
wave. In the Fraunhofer region, where the speckles have
consistently expanded, the diffraction effects ruling speckle
growth in the DFR becomes negligible, while the beam wavefront has
a radius of curvature approximately equal to the distance $z$ from
the source. Because of this curvature, two neighbor speckles of
size $\delta x =z\lambda/D$  broaden and simultaneously spread
apart at the \emph{same} rate $\vartheta_c\simeq \delta x/z \simeq
\lambda/D$. Hence, their ``paths'' do not cross any more, and each
speckle preserves its own coherence in propagation (no
``crosstalk'' between the speckles). As a consequence, the 3-D
geometrical shape of a speckle changes dramatically from a jelly
bean to a \emph{pencil}. Hence, in the Fraunhofer region, $\delta
z \rightarrow \infty$.
    \paragraph{``Full'' Fresnel region $z_c<z<z_F$}
    We have seen that speckle growth is due to two distinct
    mechanisms: diffraction, dominating in the DFR, and expansion due to wavefront curvature, which is the
    sole mechanism operating in far--field. The distance from the
    source where the two contribution becomes comparable can be
    found by equating
    $$\vartheta_d\simeq \vartheta_c \Longrightarrow \delta x \sim D
    \Longrightarrow z \simeq \frac{D^2}{\lambda}=z_F$$
    Hence, in the ``usual'', or ``full'' Fresnel region we are
    considering, both mechanisms are operating, and simple scaling arguments do not help.
    Nonetheless,  we can qualitatively say that, as the distance from the source grows, the
    wavefront becomes curved and the speckles start to spread apart.
Because of this, the diffraction patterns from two neighbor
speckles take longer to interfere, and the coherence length
becomes longer than the pure diffractive length $\delta x^2
/\lambda$. Thus, a 3D-speckle in this region cans till be viewed
a jelly beans, but sensibly elongated in the direction pointing
away from the source.

\begin{figure}[h]
\centering
  \includegraphics[width=\columnwidth]{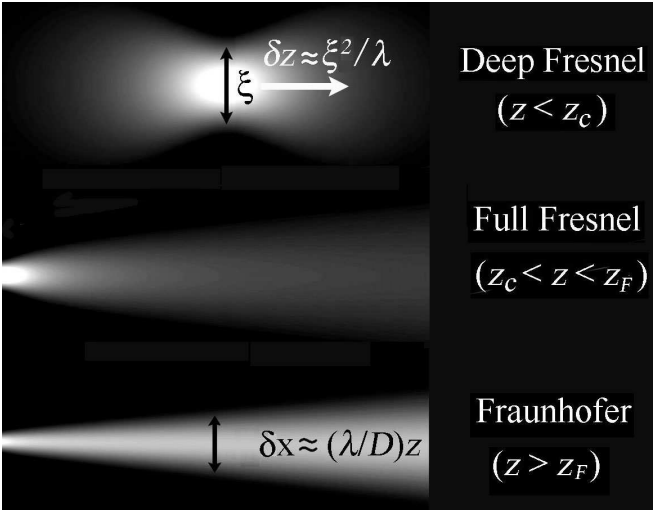}
  \caption{\label{f3} Sketch of the longitudinal coherence profile of
  speckles from an ideal monochromatic sources in the deep Fresnel, Fresnel, and Fraunhofer regions.}
\end{figure}

 A schematic view of the three regimes is shown in
Fig.~\ref{f3} (for a more quantitative description,
see\cite{Gatti}). We recall, however, that all we have said refer
to an ideal \emph{monochromatic} source, whereas the longitudinal
coherence of a polychromatic source is in any case be upper
limited by $\ell_c =c\tau_c$.

\section{Spatial coherence and imaging}
\label{s.imaging} In its simplest acceptation, imaging consist in
producing, by means of optical elements like lenses or mirrors, a
faithful copy of a planar section of an object onto another plane,
apart from a change of scale (magnification). To what extent the
copy we make can be really faithful is however limited not only by
the ``stigmatic'' properties of the imaging system (for instance
by the presence of geometrical aberrations), but also on
diffraction effects, which set the resolution limit, and therefore
the maximum useful magnification, of an imaging system. Hence, it
is not surprising that, since the seminal investigation by Ernst
Abbe, diffraction has been a fundamental tool to investigate image
formation under a microscope. Yet, spatial coherence plays a
primary role too, although this is usually marginally considered
in introductory textbooks on microscopy (with the noticeable
exception of a recent book by Mertz\cite{Mertz}). To properly
understand  how a microscope really works requires however some
basic concepts in Fourier optics and some additional results from
statistical optics.\cite{GoodmanFO,Mertz}

\paragraph{a) Angular spectrum} While investigating diffraction effects, it is usually possible
to select a ``main'' propagation direction $z$ (the optical axis),
and expanding a generic wavefront in terms of plane waves
propagating with specific components of the wave-vector
$\mathbf{k}$ along $x$ and $y$. This is done by decomposing the
amplitude $U(x,y,z)$ of a monochromatic optical field with a
\emph{partial} inverse Fourier Transform along $x, y$ as:
\begin{equation*}
    U(x, y, z) = \int \D f_x \D f_y  A(f_x,f_y,z)
    \E^{2\pi\I (f_xx+ f_yy)},
\end{equation*}
where $f_x$ and $f_y$ are called \emph{spatial frequencies} and
\begin{equation*}
    A(f_x,f_y,z) = \int \D x \D y  U(x, y, z)
    \E^{-2\pi\I (f_xx+ f_yy)}.
\end{equation*}
Spatial frequencies be given a simple geometric interpretation by
expressing the amplitude of a simple plane wave in terms of the
director cosines $(\alpha,\beta,\gamma)$ it makes with the axes
$(x,y,z)$ as $$P(x,y,z) =\exp[\I(2\pi /\lambda)(\alpha x +\beta
y)]\exp[\I(2\pi /\lambda)\gamma z].$$ Thus, across the plane $z=
0$, $\exp[2\pi \I (f_x x +f_yy)]$ may be seen as a plane wave
traveling with director cosines $\alpha =\lambda f_x$, $\beta =
\lambda f_y$. However, the director cosines are \emph{not}
independent, because $\gamma = \sqrt{1-\alpha^2-\gamma^2}.$ The
physical meaning of this relation can be grasped by observing that
$U(x,y,z)$ satisfies the Helmholtz equation
\mbox{$(\nabla^2+k^2)U(\mathbf{r})=0$}, with $k=2\pi/\lambda$.
Hence, writing $A(\alpha,\beta,z) = \int \D x \D y \, U(x, y, z)
    \E^{-\I k (\alpha x+ \beta y)}$, we have
\begin{equation*}
    \frac{\partial^2 A(\alpha, \beta,
z)}{\partial z^2} + k^2(1-\alpha^2-\beta^2)A(\alpha, \beta, z) =0
\Longrightarrow A(\alpha, \beta, z) = A(\alpha, \beta,
    0)\E^{\I k\gamma z}.
\end{equation*}
For $\alpha^2+\beta^2 \le 1$  ( $ f_x^2+f_y^2 \le \lambda^{-2}$)
$\gamma$ is \emph{real}, hence propagation just amounts to a
change of the relative phases of the components of the angular
 spectrum, because each wave travels a different distance between
constant-$z$ planes, which brings in phase delays. Conversely, for
$\alpha^2+\beta^2 > 1$ $\gamma$ is \emph{imaginary}, and $\alpha$,
$\beta$ cannot be regarded anymore as true direction cosines.
Rather, we have an \emph{evanescent} wave, whose amplitude decays
as $\exp(-2\pi|\gamma|z)$ and becomes negligible as soon as $z$
 is a few times $\lambda$. Wave propagation in free space can  then be regarded as a ``low--pass dispersive filter'', since only those
spatial  frequencies such as $f_x^2+f_y^2 \le \lambda^{-2}$ can
propagate, with a phase shift that depends however on frequency.

\paragraph{b) Fourier--Transform properties of a lens} Suppose we illuminate
with uniform amplitude $A$ a flat object, for instance a
transparency transmitting an amplitude $U(x, y) =At(x, y)$, placed
against a thin lens of focal length $f$. Then, if the object is
much smaller than the lens aperture, so that we can neglect the
effect of the finite size of the latter, the amplitude
distribution $U_f(x,y)$ in the focal plane of the lens is the
\emph{Fraunhofer diffraction pattern} of the object transmittance
$t(x, y)$, aside from a pure phase factor that does not change the
intensity.\footnote{If the finite size of the lens cannot be
neglected, $U_f(x,y)$ is actually proportional to the FT of the
product of $t(x,y)$ times the \emph{pupil} of the lens (see the
next paragraph).} The former phase factor exactly cancels out when
the object is placed at a distance $f$ \emph{before} the lens. In
other words, the front and back focal planes of a lens are related
by a FT or, as we shall say are \emph{reciprocal Fourier planes}.
 Finally, is an object is
placed before a thin lens at a distance $z_1$ then (except again
for phase factors) an \emph{image} of the object, inverted and
magnified by the ratio $M = - z_2/z_1$, forms at a distance $ z_2$
such that $z_1^{-1} +z_2^{-1}=f^{-1}$, which is of course the
simple lens law from geometrical optics. Moreover, the back focus
is exactly a Fourier plane for the object, so we can
``manipulate'' the image, for instance by  ``cutting out'' some
spatial frequencies or by selectively changing their relative
phases. This ``spatial filtering'' technique, besides being at the
roots of the whole field of optical  communication, is fully
exploited in phase--contrast microscopy. The effect of the lens
pupil is very similar, since also the lens plane is (aside from a
phase factor) a Fourier plane for the object. Hence, reducing the
lens diameter $D$ (or better, its numerical aperture $NA = D/f$)
corresponds to cut out the high-frequency Fourier components, in
fact reducing the image resolution.

\begin{figure}[h]
\centering
  \includegraphics[width=\columnwidth]{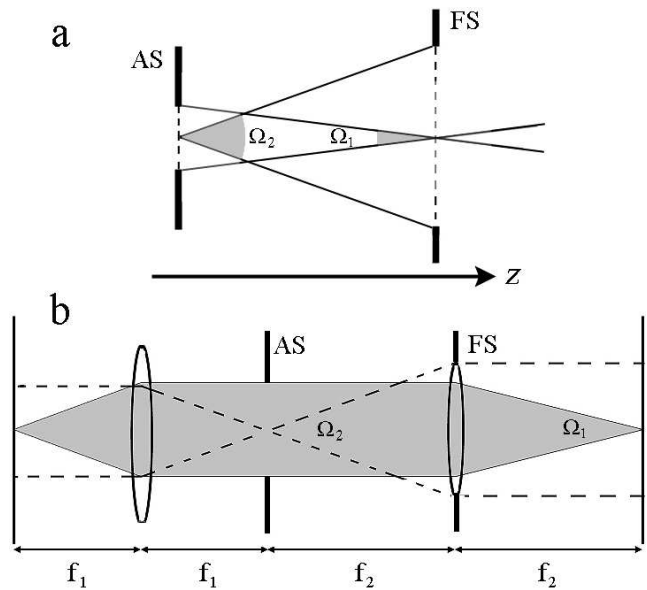}
  \caption{\label{f4} Aperture and field stops for free propagation between two apertures (a) and for a $2f_1-2f_2$ lens system (b). }
\end{figure}

\paragraph{c) Aperture and field stops} In \emph{free space}, all spatial frequencies
with $f_x^2+f_y^2 \le\lambda^{-2}$ propagate, whereas evanescent
waves die out. When an optical signal is fed through a generic
imaging system, however, there are further limitations to the
spatial frequencies that can reach the image plane, because the
finite  size of the optical components limits the angular extent
of the radiation emitted by the object that can propagate through
the system.  Crucial to the analysis of spatial coherence in an
optical system are the concepts of aperture and field stops, which
are defined as follows. Let first look at the optical system from
the image plane, and find what is the aperture which most limits
the incoming light: this is the \emph{aperture} stop AS, or simply
the ``pupil'' of the system.\footnote{Actually, in optics it is
more customary to define an ``entrance'' and an ``exit'' pupil as
the \emph{images} of the aperture stop seen through all the optics
before or, respectively, after the aperture stop. These can be
real or virtual images, depending on the location of the aperture
stop.} Now project of cone from the center of the aperture stop,
and find what is the stop that limits its angular aperture: this
is the \emph{field} stop FS. For example, Fig.~\ref{f4}a, shows
the aperture and field stops for a simple propagation between two
diaphragms, whereas in the so--called $2f_1-2f_2$ lens system
shown in Fig.~\ref{f4}b (a very convenient combination for spatial
filtering) AS is the diaphragm placed in the common focus of the
two lenses,  while FS is the pupil of the lens that limits more
the angular aperture.

\begin{figure}[h]
\centering
  \includegraphics[width=\columnwidth]{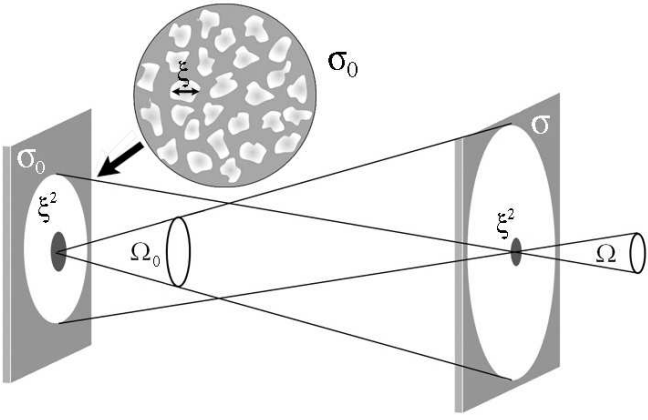}
  \caption{\label{f5}  Propagation of the spatial coherence
  for a partially--coherent source and \'{e}tendue.}
\end{figure}

\paragraph{Partially--coherent sources} According to footnote~\ref{gauss}, the
fundamental gaussian mode emitted by a laser has a far--field
angular divergence $\theta \simeq \lambda/(\pi w_0)$, where $w_0$
is the beam waist, which is the spread expected for a spatially
coherent wavefront because of diffraction. For a \emph{partially}
coherent circular source of area $\sigma_0 =\pi w_0^2$, which can
be pictured as ``speckle mosaic'' made of $N_c \sim
\sigma_0/\xi_0^2$ uncorrelated coherence regions of size $\xi_0$
(see figure~\ref{f5}), the divergence is found to be $N_c$ times
larger.\footnote{The same applies to the higher transversal modes
of a laser.}
It is however interesting to investigate how the \emph{correlation
length} changes upon propagation. We have seen that, in the deep
Fresnel region, the propagation of the spatial coherence is very
different from what predicted by the VCZ theorem for a fully
uncorrelated source. Here, however, we wish to find how a similar
source behaves in \emph{far} field, namely, in the Fraunhofer
diffraction regime. Without entering into details, which involve
rather tedious calculations, we just state the main result. In far
field, the area $\sigma$ of the source  and the correlation length
$\xi$ grow upon propagation by a distance $z$ as
\begin{equation}\label{par_coh_prop}
    \left.\begin{array}{l}
             \sigma = \dfrac{(\lambda z)^2}{\xi_0^2}\vspace{3pt}\\
             \xi^2 = \dfrac{(\lambda z)^2}{\sigma_0}\\
           \end{array}\right\}\Longrightarrow \frac{\sigma}{\xi^2}
           = \frac{\sigma_0}{\xi_0^2}.
\end{equation}
Hence, the ``expansion rate'' of the source area is determined by
the area of a coherence region and vice versa. It is therefore
useful to define a quantity with the dimensions of an area called
the \emph{\'{e}tendue}
\begin{equation}\label{etendue}
    G = \lambda^2 \frac{\sigma}{\xi^2}.
\end{equation}
which, because of Eq.~(\ref{par_coh_prop}), has the very important
property  of being \emph{conserved} upon free--space
propagation.\footnote{Note that for a fully coherent source the
\'{e}tendue attains its minimum value $G = \lambda^2$.} Moreover,
introducing as in Fig.~\ref{f5} the solid angles $\Omega_0 =
\sigma/z^2$ and $\Omega = \sigma_0/z^2$, we can also write  $ G=
\sigma_0\Omega_0 =\sigma\Omega$. Physically, the \'{e}tendue is a
``combined extension'' of the source, given by the the product of
its area in the real space times its far--field diverging angle,
which is related to the region in the Fourier space of the spatial
frequencies that propagate from $\sigma$. For a uniform source, we
can the write the total emitted power as the product $W= GL$ of
the \'{e}tendue times the radiance: since $W$ is of course fixed,
the invariance of the \'{e}tendue upon free--space propagation is
equivalent to the conservation of the source brightness.

The \'{e}tendue is however \emph{not} conserved in the presence of
limiting apertures. Suppose for instance that a fully coherent
planer wavefront of infinite lateral extent impinges on the simple
system in figure~\ref{f4}a, where the aperture stop AS limits the
source size, while the field stop FS its angular divergence: it is
easy to show that the effective \'{e}tendue is limited to $G_t=
A_s\Omega = A_f\Omega' = A_sA_f/z^2$, where $A_s$ and $A_f$ are
the areas of the aperture and field stop respectively. This is
called the \emph{throughput} of an optical system. Using the
double--diaphragm setup in figure~\ref{f4}, we can actually
\emph{increase} the effective spatial coherence of a source: this
happens whenever the solid angle subtended by FS is smaller than
than $\Omega_0$ (we loose of course some power). This also helps
to understand fiber--optic detection in DLS. A monomode optical
fiber has by definition an \'{e}tendue $G=
\sigma_f\Omega_f=\lambda^2 $ where $\sigma_f$ and $\Omega_f$ are
the area of the fiber core and its solid acceptance angle. For a
source of \'{e}tendue $G$, the maximum power fed into the fiber is
$W= W_0 \sigma_f\Omega_f/G$. It  is then easy to show that, for a
monomode fiber collecting scattered radiation, $W$ coincides with
the power scattered by the sample within one speckle.

\subsection{Microscope structure: coherence of illumination and resolution limit}
\begin{figure}[h]
\centering
  \includegraphics[width=\columnwidth]{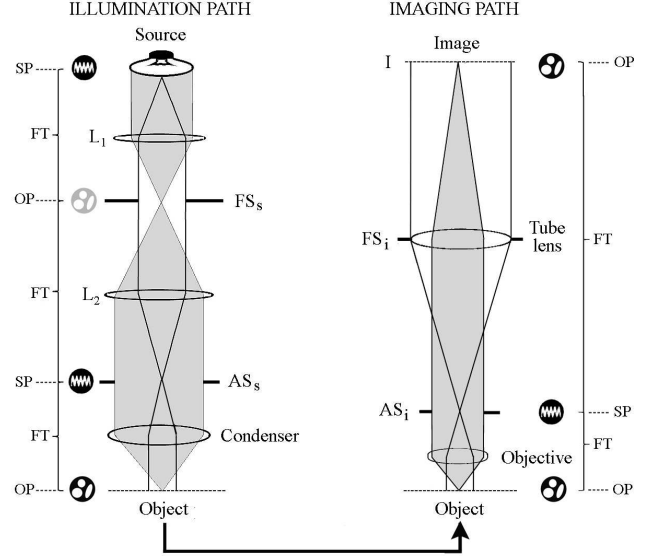}
  \caption{\label{f6} Structure of a microscope with K\"{o}hler illumination.
  The illumination path consists of the collector lenses $L_1$ and $L_2$ that generate
  an image of the illumination source on the plane of the condenser, which focuses
  the light on the object plane. In the imaging path, the transmitted light is collected by
  an infinity--corrected objective and made parallel by the tube lens.
  The two conjugate sets of planes where the illumination source and the sample are
  in focus are shown by corresponding symbols. }
\end{figure}

Fig.~\ref{f5} shows the basic structure of an optical microscope
using \emph{K\"{o}hler illumination}. This setup provides a
uniform illumination of the sample by placing the latter on the
focal plane of the condenser, which is a conjugate Fourier plane
for the illuminating lamp. As a matter of fact, in the
configuration shown in Fig.~\ref{f6} there are actually two sets
of planes where the source $S$ and the object (sample) plane are,
respectively, imaged. Set 1 is composed of the lamp filament, the
source aperture stop $\mathrm{AS}_s$ at the front focal plane of
the condenser, and the image aperture stop $\mathrm{AS}_i$ at the
back focal plane of the objective. All these planes are Fourier
planes for set 2, which comprises the field stop $\mathrm{FS}_s$
at the back focal plane of the collector lens $L_1$, the object
plane where the sample is placed, and the image plane (in visual
observation, the latter is further imaged by the eyepiece).

Understanding the reciprocal nature of these two sets of planes is
crucial to describe the way a microscope works. In particular, it
is important to stress that \emph{the size of the illumination
source and its spatial coherence properties can be controlled
independently}. The former is simply tuned by opening or closing
the diaphragm $\mathrm{AS}_s$. The field stop $\mathrm{FS}_s$
conversely controls the angular aperture of the light reaching the
sample from a given point on the source plane. Since the latter
lies on the focal plane of $L_1$, where we have the  FT of the
source, closing down $\mathrm{FS}_s$ corresponds to filtering the
spatial frequencies of $S$ and therefore to tuning the spatial
coherence properties of the source.  By increasing the condenser
aperture, the illuminating optics becomes more and more similar to
a fully incoherent source, whereas by progressively stopping it
down we approach the coherent illumination limit. From what we
have seen in the previous section, the illumination on the object
plane has then in general the form of a ``speckle mosaic'' similar
to the one sketched in figure~\ref{f5}, where the speckle size
$\xi$ is fixed by the condensed numerical aperture. In
section~\ref{s.ddm}, we shall see how novel correlation methods in
microscopy exploit this peculiar tunability of the spatial
coherence of illumination.

The degree of spatial coherence of the illumination at the sample
plane has noticeable effects on the resolving power of the
microscope. For of a telescope with an objective of radius $w$,
the determination of the resolving power is particularly simple,
because two close-by stars we may wish to resolve behave as
mutually incoherent point sources. Moreover, since the telescope
is focused ai infinity, each one of them is imaged on the focal
plane of the objective as an ``Airy disk'' (namely, the Fraunhofer
diffraction pattern of a circular aperture) of diameter $d \simeq
0.6 \lambda f/w$. A reasonable criterion for separation, suggested
by Rayleigh, is that they are ``barely resolved'' if the center of
the Airy disk of one star coincides with the first minimum of the
second one, namely, if their angular separation is larger than
$\vartheta_{min} \simeq 0.6 \lambda/w$. For microscope, however,
the problem is more complicated, first because this simple result
from Fraunhofer diffraction holds only provided that ray
propagation is paraxial, which is the case of a telescope but
surely not of a microscope; second, because we are considering
\emph{non self-luminous} objects, hence the spatial coherence of
the light generated at the object plane depend on the coherence of
the illuminating source. Consider first the situation where the
illumination is fully incoherent, which can be obtained for
instance by opening up completely the field stop $\mathrm{FS}_s$
of the condenser. If we take a look to the ``imaging path'' to the
right of figure~\ref{f6}, we can see that the spatial frequencies
of the light produced at the object plane that can reach the image
plane are  basically limited by the \emph{aperture} stop
$\mathrm{AS}_i$. Since the object plane lies very close to the
front focal plane of the objective, the maximum spatial frequency
that enters the imaging path is determined by the numerical
aperture of the latter $NA_{obj} = n\sin\vartheta$, where
$\vartheta$ is the angle subtended by $\mathrm{AS}_i$ when viewed
from the image plane, and $n$ is refractive index of the medium
the objective is immersed in (which may not be air). It is then
not hard to deduce that the Rayleigh limit is generalized by the
celebrated Abbe criterion, stating that the minimal separation
distance is:
$$ \delta \simeq 1.22\frac{\lambda}{n\sin\vartheta} = 1.22\,\frac{\lambda}{NA_{obj}},$$
For  fully \emph{coherent} illumination, however, things are quite
different, even in the paraxial approximation, and the result
depends on phase difference $\varphi$ of the illumination at the
two point sources. Indeed, one finds that the situation is
identical to the incoherent case only when the phases are in
quadrature ($\varphi = \pi/2$), whereas, when the two sources are
fully in phase ($\varphi = 0$), the two Airy disks conversely
merge into a single peak centered at $x=0$: thus, at the Rayleigh
limit, they are not resolved at all. If the sources are in
\textit{counter}-phase ($\varphi = \pi$), however, at the Rayleigh
limit they are \emph{fully} separated, hence resolution actually
doubles. Stating that coherent illumination is ``worse'' than
incoherent illumination, as often made in elementary textbooks, is
therefore incorrect. With coherent illumination, the resolution
actually depends on the specific way we illuminate the object:
whereas in a standard geometry two close-by points are usually
illuminated with the same phase, with a suitable \emph{oblique}
illumination (a technique which has often been used in microscopy)
one can obtain a counter--phase condition. Even with a standard
illumination geometry, the best resolution is not obtained by
increasing as much as possible the condenser aperture. A detailed
calculation shows indeed that it is not worth increasing the
condenser numerical aperture $NA_{con}$ to more than about
$1.5NA_{obj}$, and that in these conditions the resolving power
is\cite{BW}
\begin{equation}\label{res_pow_mic}
    \delta \simeq 1.22\,\frac{\lambda}{NA_{obj} + NA_{con}}.
\end{equation}

\section{Scattering and imaging: towards a joint venture}
We have seen how statistical optics concepts can describe both DLS
and imaging by a microscope. Yet, communication between these two
worlds has been rather limited till a few years ago. The main
reason is that the description of particle scattering necessarily
requires a full 3-D treatment of the electromagnetic problem
leading, even in the case of spherical particles, to the
complicated Lorenz-Mie solution. On the other hand, most
traditional microscopy problems can be discussed using the simpler
language of diffraction, which is basically 2-D. Recent
advancements in imaging, such as the development of confocal
microscopy and of accurate particle--tracking methods, have led to
investigate many aspects of imaging of 3-D objects, and to
reconsider the relation between scattering and microscopy.

The latter is far from being trivial. It is not easy even to state
\emph{when} we can actually see under a microscope a particle made
of a non-absorbing material and with a size much larger than the
wavelength.  If we regard them as \emph{two dimensional} sources
and just apply the basics of Fourier optics, the answer is simple:
\emph{never}. A non absorbing particle just modulates the
\emph{phase} of the illuminating radiation, and does not change
its amplitude: in other words, they are phase diffractive
elements, and the image of a phase element is again a phase
element, with no intensity contrast.\cite{GoodmanFO} Cells and
other optically transparent biological samples object are indeed
practically invisible, except at their contour boundaries, but, as
a matter of fact, large polystyrene particles \emph{can} be seen
under a microscope, even when they are right on focus. This must
have therefore to do both with the 3D nature of the particles, and
with the difference $n_p -n_s$ between the refractive indexes of
the particle and of the solvent. In fact, particles of a size $a$
such that $|n_p-n_s|a/\lambda \ll 1$ (namely, Rayleigh-Gans
scatterers) cannot be visualized at all, and the same is true for
particles scattering in the so-called ``anomalous diffraction''
regime,\cite{vdh} where reflections and refractions at the
particle/solvent interface can be neglected.\footnote{It is indeed
because of the latter that pure phase fluctuations on the object
plane yield \emph{amplitude} fluctuations when propagated to a
following plane, because of an effect similar to shadowgraphy in
geometrical optics} A detailed analysis of the visibility problem
for a generic scatterer is however still lacking. Scattering from
non-absorbing objects is in any case rather weak, whatever their
refractive index with the surrounding medium, hence a common way
to increase their visibility is ``de-focusing'', namely, focusing
the objective on a plane outside the particle. However, it is
worth noticing that, with this methods, evaluating particle size
or interparticle distances is not trivial, and may lead to serious
errors.\cite{Bechinger} In fact, quantifying how the imaging
optics collects the intensity distribution generated on a generic
plane from particles situated at various distances $z$ from it
requires a full 3D treatment of the imaging process.  In the
simplest case of a Rayleigh--Gans scatterer, one finds that the
intensity pattern consists of a central disk surrounded by a set
of concentric fringes that get the coarser the farther is the
particle from the plane $z=0$, and that a particle displacement at
constant $z_0$ amounts to a rigid translation of this fringe
pattern, similarly to what is observed in out-of-focus microscopy
observations. A full discussion of 3D imaging can be found in
Ref.\cite{Streibl,Nemoto}.

For what follows it is also useful relating the scattering
wave-vector $\mathbf{q}$ to its projection
$\textbf{q}_\shortparallel$ on the observation plane $z=0$. Since
in the paraxial approximation $q = 2k\sin(\theta/2)\simeq
k\theta$, where $\theta$ is the scattering angle, we have
\begin{equation}\label{q_plane}
    q^2 \simeq q_\shortparallel^2 \left[1+\left(
    \frac{q_\shortparallel}{2k}\right)^2\right],
\end{equation}
so that the \emph{perpendicular} component of $\mathbf{q}$ is $q_z
\simeq q_\shortparallel^2/2k$. The second term in square brackets
is of order $\theta$, so it is negligible for small scattering
angles. Notice however that, according to Eq.~(\ref{q_plane}), the
same $\mathbf{q}_\shortparallel$ vector may actually correspond,
for two distinct wavelengths, to \emph{different} $\mathbf{q}$
vectors. Nevertheless, it is not difficult to show that this
effect is small as long as the difference in wavelength $\Delta
\lambda \ll q_\shortparallel^{-1}$, which is of order
$\lambda/\theta$: hence at small collection angles, the speckle
patterns formed by different wavelengths superimpose.

\subsection{Photon Correlation Imaging (PCI)}
TRC is a very powerful method to investigate the heterogeneous and
intermittent time--dynamics of restructuring processes in gels and
glasses. However, glassy dynamics is also very heterogeneous
\emph{in space}, behaving very differently in different regions of
the sample at equal time. Photon Correlation
Imaging,\cite{Cipelletti3} a simple extension of TRC, allows to
detect these spatial heterogeneities  by means of measurements of
space and time resolved correlation functions. With respect to the
TRC scheme, the major change concerns the collection optics.
Instead of collecting the light scattered in far--field, one forms
an low-magnified \emph{image} of the scattering volume onto a
multi-pixel detector, using only the light scattered in a narrow
cone centered around a well defined scattering angle. Of course,
since the magnification $M$ is low, the scatterers themselves are
not resolved, but a speckle pattern is visible, because we are
actually collecting the light within all the depth of field of the
imaging lens, hence \emph{also} the near--field scattering from
the sample. In fact, we can \emph{tune} the size of the speckles
by adjusting the NA of the imaging lens with an iris diaphragm,
exactly as when, in heterodyne NFS, the near--field speckle
pattern is magnified using an objective. In contrast to far field
speckles that are formed by the light coming from the whole
scattering volume, however, each speckle in a PCI experiment
receives only the contribution of scatterers located in a small
volume, centered about the corresponding object point in the
sample. The linear size of this volume will be of order
$(\lambda/Md)z$, where $d$ is the diameter of the lens pupil, and
$z$ the lens-detector distance. As a result of the imaging
geometry, the fluctuations of the intensity of a given speckle are
thus related to the dynamics of a well localized, small portion of
the illuminated sample. Hence, the local dynamics can be probed by
dividing the image in ``Regions of Interest'' (RoI) which contains
a sufficient number of speckles and measuring their
time-fluctuations.

This method was developed to study slow or quasi--arrested
systems, but it works also for free particles in Brownian motion
too, provided that the speckle size is sufficiently enlarged by
stopping down the imaging lens and that a fast detector is used.
For instance, in our lab we were able to obtain very good
measurements for dilute suspensions of particles with a size of
about $50\unit{nm}$ using a fast CMOS camera. Of course, because
one measures many speckles simultaneously, the averaging process
is very fast, and very good correlation functions can be obtained
in a few seconds, but there is much more than this. Indeed, when
the particles, besides performing Brownian motion, are also moving
as a whole, the overall motion of the speckle pattern is then a
faithful reproduction of the local hydrodynamic motion within the
sample. Hence, if the speckle correlation time is sufficiently
long, the local flow velocity can be obtained by monitoring the
motion of the speckle pattern: for instance, for particles
settling under gravity, the local sedimentation velocity can be
obtained. This strategy has allowed to investigate the relation
between microscopic dynamics and large-scale restructuring in
depletion\cite{Brambilla} and biopolymer\cite{Secchi} gels.

\subsection{Differential Dynamic Microscopy (DDM)}
\label{s.ddm} 
\begin{figure}[h]
\centering
\includegraphics[width=\columnwidth]{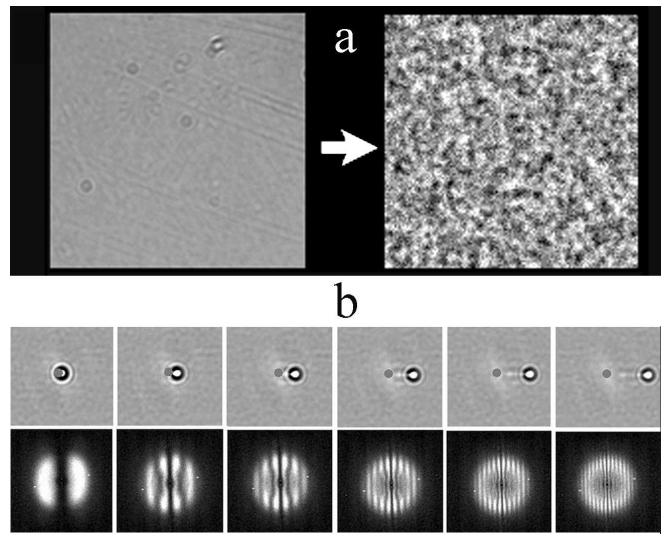}
\caption{\label{f7}  Panel A: ``Extraction'' of the speckle
pattern by image subtraction in DDM. The images refer to a
suspension of PS particles with a diameter of about $0.1\unit{\mu
m}$ at a concentration of about 0.2\%, imaged with a $0.5 NA$
objective and a stopped-down condenser. Panel B: Time-evolution of
the correlation peak (top) and of the structure factor (bottom) in
a GPV experiment.}
\end{figure}

The powers of microscopy and DLS are perfectly
combined in Differential Dynamic Microscopy (DDM), a simple but
very powerful technique  that can be set up on a standard
microscope and does not even require a coherent laser
source.\cite{Cerbino2, Cerbino3} Let us see how it works by
retracing the original steps made by R. Cerbino and V.
Trappe.\cite{Cerbino2} The image under a conventional microscope
of a suspension of particles having a size much smaller than the
wavelength is just an uniform white field with spurious
disturbances due to dust or defects in the optics, like in the
image to the left of figure~\ref{f7}a. However, taking a second
images after a time delay, and \emph{subtracting} from it the
first one, a well-defined speckle pattern appears, and gets the
sharper the longer the delay time $t$ (see figure~\ref{f7}a,
right). In fact, calling $\Delta I(x,y;t) = I(x,y;t) -I(x,t;0)$
the difference in intensity at a given point on the image plane,
one finds that the total variance
$$\sigma^2(t) = \int|\Delta I (x,y; t)|^2 \D x \D y$$ grows
with time, progressively reaching a plateau.

Why the speckles? Collecting just the field originating from the
object plane, we would not see any intensity difference between
the two frames,\footnote{Apart from effects due to particle number
fluctuations within a coherence area of the source, which should
however decrease as the inverse of the particle concentration}
but, as in PCI, we are \emph{also} collecting  the scattering in
the near field. Actually, DDM has many points in common with
near--field scattering, but with two crucial advantages. First, we
do not need at all a monochromatic source because, as discussed in
the last section, the speckle patterns generated by different
wavelengths fully superimpose at small angles. To make it clearer,
it is sufficient to observe that each spatial frequency $f_o$ of
the object behaves as a grating, diffracting in paraxial
approximation at an angle $\theta = \sin^{-1} (\lambda f)$. This
diffraction pattern generates on the image plane a set of fringes
with spatial frequency $f_i = \sin(\theta)/\lambda = f_o$ that
\emph{does not depend on} $\lambda$. Hence, each different
wavelength generates an identical interference pattern which
depends only on $f_o$, which, provided that $\Delta \lambda \ll
q_\shortparallel^{-1}$ is uniquely associated to a single
scattering wave-vector $q = 2\pi f_0$. Second,  at variance with a
standard NFS experiment with a laser source, using a microscope we
can vary the spatial coherence of the illuminating source. This
means that the deep Fresnel region where NFS is observed depends
on the numerical aperture of the condenser: in fact, if the
condenser is fully opened, no appreciable speckle pattern is
observed. What is more important, this  also amounts to change the
\emph{thickness} of the sample region which is coherently
illuminated: we have indeed seen that the speckles have a ``jelly
bean'' structure, with a longitudinal size $\delta
z\sim\xi^2/\lambda$, where $\xi$ is the transversal coherence of
the source on the object plane. By micrometrically translating the
objective, a ``$z$-scan'' through the sample can be made. The
typical longitudinal resolution is is of the order of tens of
microns, which is much larger than the resolution achievable with
a confocal microscope, but still sufficient for many purposes.

With DDM, one can in fact obtain fast measurements of the
intensity correlation function at very low angles. Recalling that
there is a one-to-one correspondence between the spatial
frequencies of the image and the scattering wave-vectors, and
using the Parseval's theorem, which states that the integral of
the square of a function is equal to the integral of the square of
its Fourier transform,\cite{GoodmanFO} the total variance can
indeed be written also as
$$\sigma^2(t) = \int|\widetilde{\Delta I} (f_x,f_y;t)|^2 \D f_x \D
f_y,$$  where $\widetilde{\Delta I} (f_x,f_y;t)=\mathscr{F}[
\Delta I(x,y;t)]$. Hence, by Fourier-transforming the image
differences, one can extract the Brownian dynamics of the
particles.\cite{Cerbino1}

\subsection{Ghost Particle Velocimetry (GPV)}
Particle Imaging Velocimetry is extensively used to monitor fluid
flow in microfluidics systems, which are becoming widespread in
academic and company research labs. Individual tracking, however,
requires particles large enough to be optically resolved, which
therefore perturb the flow over spatial scales that, in
microfluidics, may be comparable to those of the investigated
structures. This limitation can be overcome by resorting to more
sophisticated methods such as micro-scale Particle Imaging
Velocimetry ($\mu$PIV), which exploits small fluorescent tracers
that do not need to be individually resolved. In this alternative
approach, the fluid average velocity within a small region is
rather found by detecting fluorescence intensity fluctuations and
evaluating the spatial cross-correlation of two images taken at
different times with a suitable frame rate.\cite{Lindken} However,
$\mu$PIV instrumentation requires a rather expensive optical
setup, usually including a pulsed laser source synchronized with a
high resolution fast CCD camera.

As we mentioned in section~\ref{s.NFSvelo}, NFS techniques
provides a simple, efficient, and much cheeper method for tracking
fluid motion that overcomes the main limitation of standard PIV,
since particles that are smaller than the optical resolution limit
can be used. Microfluidic applications, however, require
velocimetry to be made under a microscope on microfluidic chips
that have generally a poor optical quality: feeding in an
additional laser source and setting the configuration required to
measure near-field scattering is surely inconvenient, if not
unfeasible. An alternative approach to quantitatively map fluid
flow in microfluidic devices is what we call ``Ghost Particle
Velocimetry'' (GPV), which uses the same procedures of NFS
velocimetry, but within a DDM optical scheme.\cite{Buzzaccaro}
Figure~\ref{f7}b, for instance, which refer to an experiment made
using a standard microscope and white light, shows that two basic
strategies for extracting the local fluid velocity discussed in
Section \ref{NearField} can be used with no relevant change in a
DDM configuration.

At variance with a standard NFS experiment, however, the depth of
the region probed in GPV is extremely limited, because of the very
small size of coherence area of the illumination source: in fact,
it is much smaller than the depth of focus of the objective, so
that mapping of the velocity field can be done by focusing the
objective on the object plane itself. In a microfluidic geometry,
this allows to simultaneously obtain, for instance, a detailed
image of the channel. GPV also allows for an appreciable
resolution along the optical axis, yielding 2D sections of the
flow pattern separated by a few tens of micrometers. What is
really interesting, however, is that the size of the particles
used as tracers does not matter, as long as they scatters
sufficiently strong (remember indeed that, even for scatterers
with a size $a\ll \lambda$, the near--field speckle size cannot be
smaller than about $\lambda$, whereas their size on the image
plane is just fixed by the NA of the objective). In fact, using
GPV one can perform a detailed analysis of hydrodynamic flow using
as tracers nanometric ``ghost'' particles that are far smaller
than the microscope resolution limit.\cite{Buzzaccaro}

PCI, DDM, and GPV are just some examples of how a careful
application of statistical optics concepts can help in devising
novel powerful optical methods that bring together scattering and
imaging.  In fact, these techniques, and DDM in particular, are
deeply related to other methods that fully exploit coherence
effects, such as Digital Holography and Optical Tomography. It is
therefore highly probable that in the next future these new
fascinating approaches will gain more importance in the
investigation of colloidal systems.

\end{document}